\documentclass[prb,aps,superscriptaddress,longbibliography, onecolumn]{revtex4-2}

\pdfoutput=1
\usepackage[utf8]{inputenc}
\usepackage[english]{babel}
\usepackage[T1]{fontenc}
\usepackage{amsmath}
\usepackage{amssymb}
\usepackage{physics}
\usepackage{hyperref}
\usepackage{setspace}
\usepackage{braket}
\usepackage{bbm}
\usepackage{graphicx}
\usepackage{amsthm}
\usepackage{appendix}
\usepackage{natbib}

\usepackage[dvipsnames]{xcolor}
\definecolor{orange}{RGB}{255,127,0}

\begin{abstract}

We study a one-dimensional frustrated spin chain, which combines cluster-Ising  and anisotropic next-nearest neighbor Ising 
models. We first offer a historical perspective that justifies the studied model. Then we study in detail the two quantum phases and prove that they are separated by a first order quantum phase transition. On one side, the ground state corresponds to a ferromagnetic phase, shows the presence of macroscopic cat states, and a small gap that closes in the thermodynamic limit. On the other phase, competing interactions avoid the establishment of a topological phase, though it conserves large incommensurate  quantum correlations. We prove it is fundamentally distinct from a simple paramagnet, and we name it an incommensurate phase. This is a gapped phase, which gap does not close in the thermodynamic limit. While in the ferromagnetic phase there are two dominant Schmidt coefficients, in the incommensurate phase there are four. This corresponds to four distinct bipartite entanglement channels contributing substantially to the ground state. Finally, we discuss the utility of the macroscopic cat states for quantum metrology applications and the experimental feasibility of the system.
\end{abstract}
\begin{document}

\title{Emergence of a Macroscopic Cat State and Multi-Channel Entanglement in a Frustrated Cluster Spin Chain}

\author{Mohit Lal Bera}
\email{molalbe@uv.es}
\affiliation{Departamento de F\'{i}sica Te\'{o}rica and IFIC, Universidad de Valencia-CSIC, 46100 Burjassot (Valencia), Spain}

\author{Andreu Angl\'{e}s-Castillo}
\email{aangcas@upv.es}
\affiliation{Departamento de F\'{i}sica Te\'{o}rica and IFIC, Universidad de Valencia-CSIC, 46100 Burjassot (Valencia), Spain}
\affiliation{Departament d'Inform\`{a}tica de Sistemes i Computadors, Universitat Polit\`{e}cnica de Val\`{e}ncia, Val\`{e}ncia, Spain}

\author{Alberto Acevedo Mel\'{e}ndez}
\affiliation{Departamento de Matem\'{a}ticas, F\'{i}sica y Ciencias Tecnol\'{o}gicas, Universidad CEU Cardenal Herrera, Val\'{e}ncia, Spain}
\affiliation{IUMPA - Instituto Universitario de Matemática Pura y Aplicada, Universitat Politècnica de València, E-46022 València, Spain}

\author{Rafael G\'{o}mez-Lurbe}
\affiliation{Departamento de F\'{i}sica Te\'{o}rica and IFIC, Universidad de Valencia-CSIC, 46100 Burjassot (Valencia), Spain}

\author{Luca Ion}
\affiliation{Departamento de F\'{i}sica Te\'{o}rica and IFIC, Universidad de Valencia-CSIC, 46100 Burjassot (Valencia), Spain}

\author{Rodrigo M. Sanz}
\affiliation{Departament d'Inform\`{a}tica de Sistemes i Computadors, Universitat Polit\`{e}cnica de Val\`{e}ncia, Val\`{e}ncia, Spain}

\author{Somayeh Mehrabankar}
\affiliation{Queensland Quantum and Advanced Technologies Research Institute, Griffith University, Yuggera Country, Brisbane, QLD 4111, Australia}

\author{Tanmoy Pandit}
\email{tanmoy.pandit@vtt.fi}
\affiliation{VTT Technical Research Centre of Finland, Tietotie~3, Espoo, Finland}

\author{Carmen G. Almud\'{e}ver}
\affiliation{Departament d'Inform\`{a}tica de Sistemes i Computadors, Universitat Polit\`{e}cnica de Val\`{e}ncia, Val\`{e}ncia, Spain}

\author{Armando P\'{e}rez}
\affiliation{Departamento de F\'{i}sica Te\'{o}rica and IFIC, Universidad de Valencia-CSIC, 46100 Burjassot (Valencia), Spain}

\author{Miguel Angel Garcia-March}
\email{Corresponding author: garciamarch@upv.es}
\affiliation{IUMPA - Instituto Universitario de Matemática Pura y Aplicada, Universitat Politècnica de València, E-46022 València, Spain}

\maketitle




\section{Introduction}

\paragraph*{Historical background}
The description of the out-of-equilibrium open dynamics of quantum systems was and still is a crucial problem of quantum mechanics. The  Gorini-Kossakowski-Sudarshan-Lindblad (GKSL) equation, also known as the master equation in Lindblad form,  is one of the general forms of the Markovian master equations describing open quantum systems. It was a celebrated result that is central to this body of research~\cite{1976Lindblad} and a generalization of Schr{\"o}dinger equation. During the 1970s there were many efforts devoted to this problem. In the following decades, all this work was synthesized in what we currently know  as the theory of open quantum systems~\cite{1993Carmichael,2002breuer,2012Weiss}. In this context of research during the 1970s and years before one finds a remarkable paper from  Glauber, where he studied out-of-equilibrium stochastic dynamics of the one-dimensional Ising Model~\cite{Glauber1963}. It may also help to understand the context for this paper to remind that Prof. Glauber joined with only eighteen years project Manhattan and witnessed firsthand the development of Monte Carlo methods for physics problems~\cite{2023Latorre}.    

In  Glauber's 1963 paper~\cite{Glauber1963}, he assumed that a (classical) Ising chain with constant interaction parameter $J$ -- and no magnetic external field --  is coupled to a thermal reservoir at inverse temperature $\beta$. Each Ising variable makes transitions randomly between two values,  caused by the heat reservoir. He assumed also that the "transition probabilities of the individual spins are assumed to depend on the momentary values of the neighboring spins as well as on the influence of the heat bath". Then, following stochastic methods~\cite{2004Gardiner} he postulated a Markovian jump equation, also known as Pauli Master equation~\cite{2002breuer}.  Here, the probability of every configuration of spins evolves in time due to the balance of jumps from other configurations differing only in one spin and with certain rates. This is a Markov process of $N$ discrete random variables with a continuous time variable as argument. To determine the rates, he accounted for what is expected to occur for such chain (to favour ferromagnetic or antiferromagnetic states), he also assumed detailed balance condition and that the final equilibrium state is that of the thermal Gibbs equilibrium given by the Hamiltonian of the Ising system. Simple as it is, it nevertheless was crucial for developing Glauber Monte Carlo method (also known as Kinetic Monte Carlo, dynamic Monte Carlo or Gillespie Method) which is crucial to e.g. Nuclear physics~\cite{2007Miller} and allows one for classical numerical simulations of Ising models in one-, two- or any dimension. Remarkably, this was used  to calculate the dynamic $z$ relaxation critical exponent~\cite{1980Haake} - which calculation is used to compare classical and quantum random numbers~\cite{2024Cirauqui}. 

Next step is crucial for this paper: the resulting Master equation can be regarded as an imaginary-time Schr\"{o}dinger equation, with an  effective Hamiltonian, that we call  Glauber-Ising Hamiltonian -- see Eq.~\eqref{eq:H_gammadelta} below). 
This  was noted by  Felderhof and co-workers~\cite{1971Felderhof,1972Hilhorst,1977Siggia} and it was an active topic of research at the beginning  of the 1970s. In particular, Felderhof showed~\cite{1971Felderhof} that the Glauber-Ising Hamiltonian (when $\delta=0$ in Eq.~\eqref{eq:H_gammadelta} below) can be diagonalized with a conventional Jordan-Wigner transformation, followed by Fourier transform and Bogoliubov rotation.  Then, the  Glauber-Ising Hamiltonian was the generator of the out-of-equilibrium dissipative dynamics of the spin chain, led by the initial jump equation. We remind that this was developed around 1970, and before Lindblad Master equation was introduced. So, in what way the open dynamics can be understood from this  Hamiltonian?   The description was over-simplistic, in what regarded the environment and interaction of the system with the environment, and a deeper description required including assumptions over the quantum dynamics of the open system, defining the operators and states associated to it, and working carefully within the interaction picture. Nevertheless, see e.g. section 3.3.2 of~\cite{2002breuer}, this can be obtained from the  diagonal part of the full open quantum dynamics.  
 In other words, the  Hamiltonian accounted well for the diagonal elements of the reduced density matrix of the system, which is expected to reach a thermal equilibrium with an environment at equilibrium, described by its reduced density matrix. One important example that tackled this question was offered in~\cite{Augusiak2010}, where they wrote a quantum Lindbland equation for the dynamics of a open system led by a generalization of the  Glauber-Ising Hamiltonian, termed as the quantum Kinetic Ising model.   

\paragraph*{Motivation: the quantum kinetic Ising model as a candidate for computation,  related Hamiltonians,  and Schrödinger cat states} 

Our first motivation is to perform a quantum many-body  analysis of this system which will permit one  to take advantage of the Glauber-Ising Hamiltonian  in applications for optimization, i.e., to encode information in the coefficients, as is done with general Ising Hamiltonians in quantum adiabatic computing.
For that end,  one needs a deep understanding of the quantum phases associated to this Hamiltonian, since the performance of the above mentioned algorithms might depend on the particular quantum phase where the dynamics takes place. This is analogous and well known in other system, like Ising model\cite{QA_Qop,Sachdev2011}. 
Thus,  we ask  whether it exhibits quantum phases and  critical points as a function of the control parameter $\delta$ and its scaling to the thermodynamic limit with the  system size $N$. Importantly, we keep  the other parameter, $\gamma$ in Eq.~\eqref{eq:H_gammadelta} below, equal to one. In the classical Glauber Hamiltonian this corresponds to the case of zero temperature. In our numerical calculations we found that decreasing $\gamma$ only makes the quantum phase  transitions less pronounced. Actually, decreasing $\gamma$, which in the classical limit corresponds to increasing $T$, has the expected effect of increasing temperature in the reservoir. 

A second motivation arises by contextualizing Hamiltonian~\eqref{eq:H_gammadelta} within a family of Hamiltonians that include cluster three-body terms. In this context one notices that Hamiltonian~\eqref{eq:H_expanded} includes the two conventional Ising terms: nearest-neighbor (NN or $ZZ\mathbb{I}$) and the transverse flipping field, $\sigma^x_i$ ($\mathbb{I}X\mathbb{I}$). It also contains a cluster three-body  term $\sigma^z_{i-1}\sigma^x_{i}\sigma^z_{i+1}$, as in the cluster-Ising~Hamiltonians~\cite{2011Smacchia} ($ZXZ$). These models can be realized in triangular lattices for ultracold atoms~\cite{2010Becker}, show strongly correlated phases with topological physics~\cite{2026Yu}, and are candidates for several quantum applications, particularly to test quantum convolutional neural networks~\cite{2022Herrmann,2022Lib} or, as is one of our motivations, to implement quantum annealing protocols~\cite{2017Tanaka}. 
Finally, there is an additional term,  the next-nearest-neighbor (NNN) $\sigma^z_{i-1}\sigma^z_{i+1}$ ($Z\mathbb{I}Z$). This term appears in the axial/anisotropic next-nearest neighbor Ising (ANNNI) model~\cite{1984Selke,Selke1988}, which provides theoretical explanation for multiple systems~\cite{Elliott1961,Fisher1980,Bak1982,1984Price}.  
Further generalizations are possible, assuming any value of these coefficients or other terms, e.g. for restoring full symmetry one may introduce $XZX$, $X\mathbb{I}X$, $\mathbb{I}Z\mathbb{I}$ and break afterwards symmetry in various ways (following a scheme which mimics the fact that Heisenberg model is a generalization of Ising model - see for example~\cite{2004Pachos,2009skrovseth,2011Li}). 

A third motivation is given by the results offered by the theoretical phases and quantum phase transition of this system, particularly the appearance of macroscopic Schrödinger cat states.  One may name as a {\it microscopic} Schrödinger cat state to  a coherent superposition of two distinct quantum states. They challenge classical determinism by showing that quantum objects need not occupy definite states prior to measurement~\cite{Leggett1980, Monroe1996}. They are central even for  foundational questions about physical reality~\cite{Zurek2003, Schlosshauer2005,Wineland2013} and  are powerful resources for quantum metrology or for quantum information processing~\cite{Giovannetti2004,Raimond2001,Nielsen2012, Degen2017}. Experimentally, they have been realized in e.g. trapped ions, superconducting circuits, optical fields, and even massive mechanical oscillators~\cite{Monroe1996,Hofheinz2009,Gisin2006,OConnell2010}. This work was recognized by the 2012 Nobel Prize to Serge Haroche and David Wineland for measuring and manipulating individual quantum systems~\cite{NobelPrize2012}.

Even more challenging are  {\it macroscopic} Schr\"{o}dinger cat states, which involve a coherent superposition of two classically distinct configurations of an entire many‑body system~\cite{Leggett2002, NobelPrize2025}. The defining characteristic of macroscopic cat states is the collective participation of $N$ particles in a single degree of freedom, with coherence protected by an exponentially small energy gap that leads to a degenerate ground‑state manifold in the thermodynamic limit~\cite{Miyashita2001, Sachdev2011}. Philosophically, these states probe whether quantum superposition persists at scales where we normally experience a classical world~\citealp{Leggett2008}. Experimentally, they offer precision tests of quantum mechanics in uncharted regimes, and technologically, they enable quantum sensors with Heisenberg‑limited sensitivity, fault‑tolerant quantum computing, and quantum simulation of many‑body physics~\cite{Hyllus2012, Fowler2012,Bloch2012, Giovannetti2004} . The direct realization of such a macroscopic superposition was honored with the 2025 Nobel Prize to John Clarke, Michel Devoret and John Martinis for the discovery of macroscopic quantum tunneling and energy quantization in an electrical circuit~\cite{NobelPrize2025}. Collectively, these achievements demonstrate that quantum coherence is not limited to the microscopic world but can be engineered and observed in systems large enough to be seen, with profound implications for both fundamental physics and emerging quantum technologies~\cite{Zurek2003, Schlosshauer2005}.

\paragraph*{Description of the calculations and organization of the manuscript}

To analyze the system numerically, we combine exact diagonalization and matrix-product-state (DMRG) calculations. 
We  evaluate: (i)   spin-spin correlations, and (ii) the static structure factor (which is  the Fourier transform of the spin-spin correlations); The later two quantities signal  the quantum phase transition (QPT) from ferromagnetic to incommensurate order, but  they just contain two-body correlations. Thus, to complete the characterization of the ground state we compute higher-order correlations. In particular, we compute (iii) the fourth-order Binder cumulant. After this,  we calculate (iv) the spectral gap, which is crucial to understand the nature of the QPT. 
Once the appearance of the QPT is established and characterized we dig deeper in the utility of the macroscopic quantum coherence found in this many-body system. To this end we show (v) that the system shows in some regimes the macroscopic superposition of two ferromagnetic states, i.e. the formation of Schrödinger-cat states. Then,  we quantify further the macroscopic coherence with (vi) the quantum Fisher Information. We finally calculate (vii) the entanglement entropy and (viii) the entanglement structure, with computation of the full distribution of Schmidt eigenvalues. 

Our study reveals a first-order QPT at $\delta\simeq 0$. The  QPT  occurs between a gapless (in the thermodynamic limit) ferromagnetic phase and a gapped inconmensurate phase. 
Our main contribution is  the relevance of our results to possible quantum sensing technologies. We find   macroscopic cat-states appearing  from the bare Hamiltonian without elaborated engineering strategies, like for example quantum control.

The paper is organized as follows. In Section~\ref{sec:model} we introduce the model Hamiltonian and we discuss briefly apriori expected role of each term together with some evident symmetries.  
Section~\ref{sec:Results} presents the results related to characterization of QPT and thus corresponding to items (i) to (iv) above in subsections~\ref{subsec:corr_zz} to \ref{subsec:spectralgap}, and results related to Schr\"{o}dinger cat states, quantum Fisher information, and characterization of entanglement (i.e. items (v) to (vii) above) in subsections~\ref{subsec:cat} to \ref{subsec:entangstruc}. 
We offer a brief interpretation of the role played by all terms in the Hamiltonian in Sec.~\ref{sec:suu} and  a thorough discussion about the results, with implications and outlook in Sec.~\ref{sec:Discussion}. Importantly, we also comment in this section about the experimental feasibility of the results presented.


 \section{Model}
\label{sec:model}

The Glauber-Ising Hamiltonian~\cite{Glauber1963} reads
\begin{equation}
\label{eq:H_gammadelta}
H(\gamma,\delta)
= -\Gamma \sum_{i=1}^{N}
\Bigl\{
\bigl[A(\gamma,\delta)-B(\gamma,\delta)\,\sigma^z_{i-1}\sigma^z_{i+1}\bigr]\sigma^x_{i}
-\bigl(1+\delta\,\sigma^z_{i-1}\sigma^z_{i+1}\bigr)
\Bigl[1-\tfrac{\gamma}{2}\,\sigma^z_{i}(\sigma^z_{i-1}+\sigma^z_{i+1})\Bigr]
\Bigr\},
\end{equation}
where $N$ is the total number of spin sites.  In Eq.~\eqref{eq:H_gammadelta}, we have defined
\begin{equation}
\label{eq:A_B_defs}
A(\gamma,\delta)
=\frac{(1+\delta)\,\gamma^2}{2\bigl(1-\sqrt{1-\gamma^2}\bigr)}-\delta, \ \
B(\gamma,\delta)=1-A(\gamma,\delta).
\end{equation}
 Here $\gamma=\tanh(2\beta J)\in[0,1]$ parametrizes the bath temperature ($\gamma=0$ is infinite temperature and $\gamma=1$ zero temperature), $\delta\in[-1,1]$ tunes a three-spin terms, and $\Gamma>0$ sets the overall rate, which determines the timescale for evolution.  The way the coefficients are introduced, via $\gamma$ and $\delta$ in Eq.~\eqref{eq:H_gammadelta} is  motivated by the fact that its diagonal terms resemble the out-of-equilibrium dynamics introduced by Glauber~\cite{Glauber1963}. For the Glauber Master (jump) equation, $\delta=0$. Generalizations of this for $\delta\ne0$ were introduced later by~\cite{Haake1980,1979Kimball,1979Deker}. In this work, we are concerned with static properties of the model, therefore we assume $\Gamma=1$ without loss of generality.
 
 For a better understanding of the different roles played by the terms appearing in Eq.~(\ref{eq:H_gammadelta}), it is convenient to expand it as 
\begin{align}
\label{eq:H_expanded}
    H(\gamma \ \delta)
=  \sum_{i=1}^{N}[
 - A(\gamma, \delta) \sigma^x_{\ i} + B(\gamma, \delta)\ \sigma^z_{\ i-1} \sigma^x_{\ i}\sigma^z_{\ i+1} +1 + \delta\ \sigma^z_{\ i-1}\sigma^z_{\ i+1}-  \frac{ \gamma }{2} (\delta +1)\sigma^z_{\ i-1} \sigma^z_{\ i} - \frac{ \gamma }{2}(\delta +1)\sigma^z_{\ i}\sigma^z_{\ i+1} ].
\end{align}
To complete the description of the model, one needs to establish how boundary conditions are managed. Throughout this manuscript we consider two conventional boundary conditions, open (OBC) and periodic (PBC). Notice Glauber (and Felderhof in~\cite{1971Felderhof}) considered the latter.

%



 The combination of the conventional Ising terms -- transverse field ($\sigma^x_i$) and  NN  interactions ($\sigma^z_i\sigma^z_{i+1}$)-- with the new NNN interactions ($\sigma^z_{i-1}\sigma^z_{i+1}$) and the cluster  three-body  term ($\sigma^z_{i-1}\sigma^x_i\sigma^z_{i+1}$)
 creates a rich interaction landscape  where the tendencies towards conventional magnetic order and frustration  are in direct competition. 
 The NN Ising and NNN terms establish the primary ordering mechanism. The NN promotes ferromagnetic (FM) order both for a coupling $\delta < 0$ or  $\delta > 0$, because in every case the coefficient in front of it is negative. On the other hand, the NNN term similarly promotes FM order for $\delta < 0$ but AFM order for   $\delta > 0$. The transverse field induces quantum fluctuations that oppose any $z$-directional magnetic order. 
The crucial element of our model is the cluster  three-body term. It is known, in other contexts, to be responsible for complicated frustration~\cite{2011Smacchia,2026Tan}. 


We found that the ground state of Hamiltonian Eq.~(\ref{eq:H_expanded})  for finite chains is non-degenerate for all values of the $\delta$ parameter, except for $\delta=0$ where one encounters a double degeneracy. This particular case avoids a direct comparison with the rest of ground states, as it is represented by a two-dimensional subspace, rather than a single state. We remark that  for $\delta=0$  the Hamiltonian is diagonalized analytically~\cite{1971Felderhof}. 
We also find that the gap closes in the thermodynamic limit $N\to\infty$ for $\delta < 0$. 

Evident symmetry appears noticing that the Hamiltonian is invariant under the transformation induced by the $\mathbb{Z}_{2}$  ``flipping'' operator 
\[
F=F^{\dagger}=\otimes_{i=1}^{N}\sigma_{i}^{x}.
\]
Under the action of $F$, every $\sigma_{i}^{z}$ is transformed to
$-\sigma_{i}^{z}$. Therefore, for non-degenerate states (such as
the ground state when $\delta\neq0)$ one necessarily has 
\[
\langle\sigma_{i}^{z}\rangle=0,
\]
and this will have consequences when calculating some quantities, as for example correlation functions or the Quantum Fisher information.

Below we provide numerical density matrix renomarlization group (DMRG) computations  performed using ITensor.jl~\cite{Fishman2022} for chain lengths $N = 20, 30$, and $40$.  The results for $N = 20$ were benchmarked against exact diagonalization (ED) using the QuSpin package~\cite{Weinberg2019}; the relative energy difference between DMRG and ED was below $10^{-9}$ in all cases. 

\section{Results}
\label{sec:Results}

\subsection{Spin-Spin Correlation Function}
\label{subsec:corr_zz}

\begin{figure}[ht]
    \centering
    \includegraphics[width=1.0\textwidth]{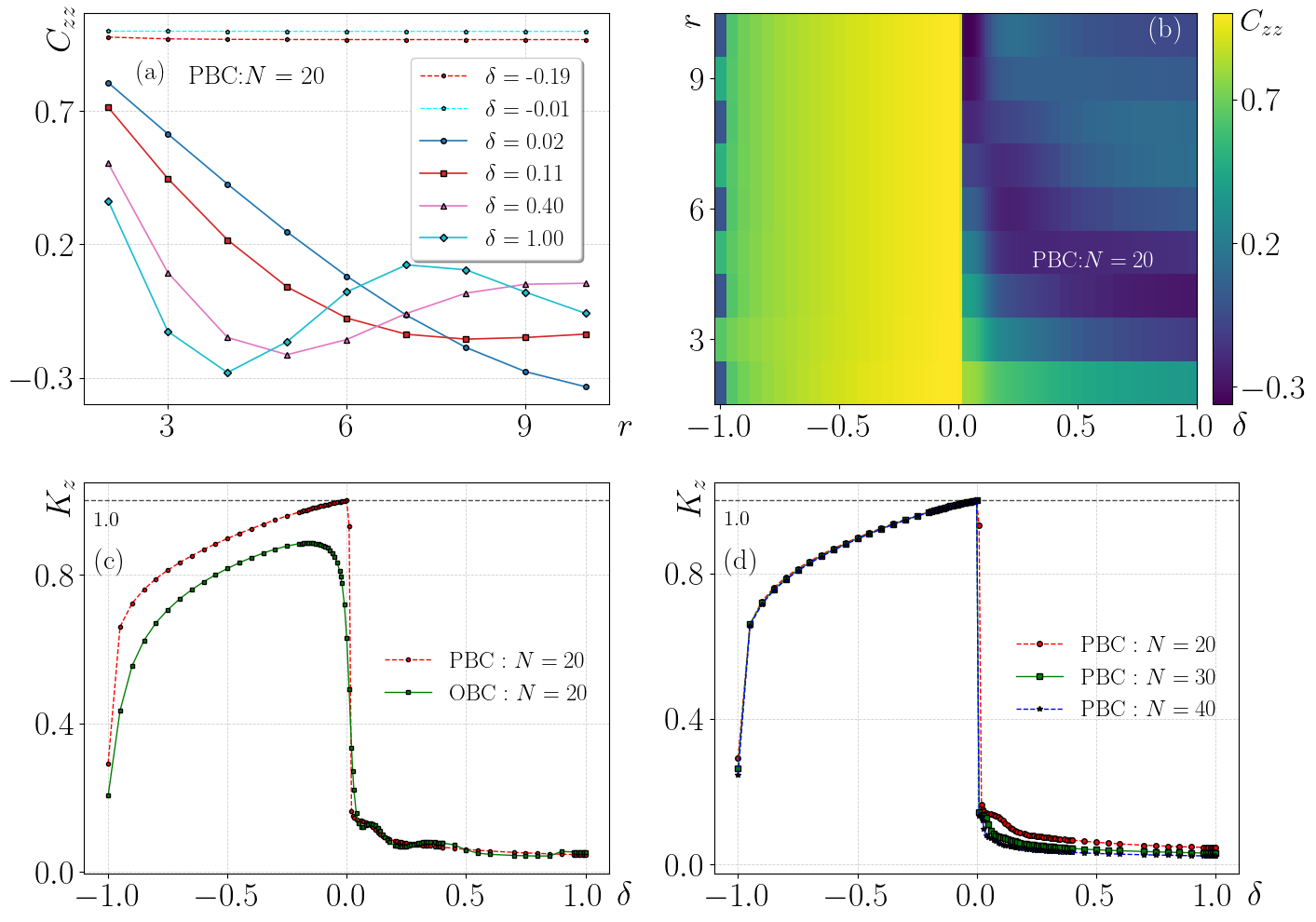}
    \caption{
    \textbf{(a)} Spin-spin correlation function $C_{zz}(r)$ versus distance $r$ for selected $\delta$ in a $N=20$ chain (PBC). For $\delta < 0$, $C_{zz}(r)$ saturates to a distance-independent constant at large $r$, confirming long-range ferromagnetic order despite $\langle \sigma^z_k\rangle = 0$ in finite systems. The saturation value increases monotonically from $\approx 0.85$ at $\delta = -0.6$ to $\approx 0.97$ at $\delta = -0.2$, approaching unity as $\delta \to 0^-$ (see main text for clarification on nomenclature). For $\delta > 0$, $C_{zz}(r)$ decays rapidly, exhibiting oscillatory components and variable decay rates reflecting incommensurate correlations distinct from simple paramagnetic decay.
    \textbf{(b)} Heatmap of $C_{zz}(r)$ versus $\delta$ (horizontal) and $r$ (vertical). The change across $\delta = 0$ is dramatic: for $\delta < 0$, strong correlations (red) persist to all distances; for $\delta > 0$, correlations decay rapidly to near-zero (blue), with oscillatory patterns appearing as alternating bands for intermediate positive $\delta$.
    \textbf{(c)} Integrated correlation $K_z$ comparing  OBC and PBC cases. Periodic boundaries enhance correlations in the ordered phase ($\delta < 0$), while open boundaries suppress them. The sharp transition at $\delta \approx 0$ confirms the bulk nature of the phase change.
    \textbf{(d)} Finite-size scaling of $K_z$ versus $\delta$ for PBC chains with $N=20$, $30$, $40$. For $\delta < 0$, rapid convergence with system size confirms true long-range order; for $\delta > 0$, systematic decrease with increasing $N$ indicates short-range correlations that vanish in the thermodynamic limit. The crossing point estimates $\delta_c \approx 0$.
    }
    \label{fig:zz_cor}
\end{figure}

We compute the spatial two-body spin-spin correlation function. This function will help characterize the quantum phases. This is defined as
\begin{align}
C_{zz}(r) = \langle \sigma^z_k \sigma^z_{k+r} \rangle - \langle \sigma^z_k \rangle \langle \sigma^z_{k+r} \rangle,
\label{eq:Czz_def}
\end{align}
where $r$ denotes lattice separation. For OBC we use $k=1$ and  range $r=1,\dots,N-1$. For PBC, translational invariance ensures independence of reference site $k$, so we simply omit $k$-dependence in the equation.  Remember that, as discussed above, $\langle \sigma^z_k \rangle = 0$.

Before discussing the results obtained, we notice that there are two  cases which usually are found when analyzing  this quantity: A) Long-range ferro- or antiferromagnetic order, where the spatial two-body spin-spin correlation function tends to some positive constant $m^2$, i.e., $\lim_{r \to \infty} C_{zz}(r) = m^2 > 0$~\cite{Sachdev2011, White1996,Kramers1941,Onsager1944}. This long-range ferromagnetic order is thus associated to spontaneous $\mathbb{Z}_2$ symmetry breaking. As with a quantum particle in a double well, this symmetry can be restored via an analog of quantum tunneling (associated to the small gap between the two quasi-degenerate states) in all finite systems. But the quasi-degeneracy becomes full degeneracy in the thermodynamic limit and spontaneous symmetry breaking cannot be avoided in this limit. We will see signatures of this in our system. 
B) Disordered phase, when $C_{zz}(r)$ decays to zero \cite{Pfeuty1970, Suzuki1976}. These two cases, A and B will serve us as expected cases to which we compare our numerical results.


Let us discuss now our results for $\delta < 0$. Here, the numerical calculations provide robust evidence for long-range ferromagnetic correlations. As shown in Fig.~\ref{fig:zz_cor}(a), $|C_{zz}(r)|$ saturates to a distance-independent constant at the largest separations ($r = N/2 = 10$). The saturation value increases monotonically as $\delta$ approaches zero from smaller values (which from here on we term as $\delta \to 0^-$). Particularly, the values obtained range from $\approx 0.85$ at $\delta = -0.6$ to $\approx 0.97$ at $\delta = -0.2$ (Fig.~\ref{fig:zz_cor}a). Further, as $\delta \to 0^-$, $C_{zz}(r) \to 1$, indicating near-perfect ferromagnetic alignment~\cite{Kramers1941,Onsager1944, Baxter1985}. This monotonic increase reflects progressive suppression of quantum fluctuations approaching the critical point from the ordered side~\cite{Sachdev2011, Dutta2015}. The persistence of long-range correlations confirms a ferromagnetic phase matching the scenario described above. As stated before, the $\mathbb{Z}_2$ symmetry is not broken in these finite systems, as  $\langle \sigma^z_k \rangle = 0$ for all range of parameters. Nevertheless, the results on $C_{zz}(r)$ show the presence of this ferromagnetic phase. One would expect that this  symmetry  be  broken in the thermodynamic limit as the infinite barrier in the infinite system is set, and the full degeneracy is established~\cite{Vidal2003,2010Mussardo,Tasaki2020}. To discuss this aspect, we will study the gap as a function of $N$ in subsection~\ref{subsec:spectralgap}. Additionally,  we note that, in this phase,  quantum fluctuations introduced via the cluster term are insufficient to destroy magnetic order (we discuss this more lengthy later, in subsection~\ref{sec:suu}). 

Our calculations for $\delta > 0$ show that the system exhibits fundamentally different behavior in this regime. Here, $C_{zz}(r)$ As shown in Fig.~\ref{fig:zz_cor}(a) and quantified in the heatmap of Fig.~\ref{fig:zz_cor}(b), the correlation function drops to negligibly small values within a few lattice spacings, confirming the disordered nature of the ground state. This behavior is driven by strong quantum fluctuations induced by the cluster three-body term (see discussion in subsection~\ref{sec:suu}), that destroy magnetic alignment (Fig.~\ref{fig:zz_cor}a,b)~\cite{Pfeuty1970, Suzuki1976}.

Importantly, this disordered phase is {\it not} a simple paramagnet. For $0.2 \lesssim \delta \lesssim 1$, $C_{zz}(r)$ displays clear oscillations, indicating competing length scales and incipient incommensurate correlations arising from frustration~\cite{Papanikolaou2012, White1996}. Though the results for $\delta=0.2$ are not displayed in Fig.~\ref{fig:zz_cor}(a), our numerical results show that $C_{zz}(r)$ starts to have oscillations around this value. The decay envelope varies significantly with $\delta$: relatively slow just above zero (consistent with proximity to criticality)~\cite{Sachdev2011}, becoming increasingly rapid as $\delta$ increases. The heatmap in Fig.~\ref{fig:zz_cor}(b) visualizes this behavior: for $\delta < 0$, strong correlations (red) persist at all distances; for $\delta > 0$, correlations rapidly decay to near-zero (blue) with oscillatory patterns appearing as alternating bands.

Results for OBC are similar qualitatively to those for PBC. Nevertheless they present some differences. To  compare among PBC and OBC let us use the integrated measure
\begin{align}
K_z = \frac{1}{N}\sum_{r=1}^{N-1} C_{zz}(r),
\label{eq:Kz_def}
\end{align}
which sums correlations over all distances~\cite{White1996, Papanikolaou2012}. Figure~\ref{fig:zz_cor}(c) compares $K_z$ for OBC and PBC with $N=20$. For $\delta < 0$, $K_z$ under PBC is substantially larger than under OBC, reflecting enhancement of long-range correlations by periodic boundaries that allow coherent propagation across the chain~\cite{Alcaraz1987, Lin1990}. For $\delta > 0$, both boundary conditions yield small $K_z$, though PBC values remain slightly elevated. 

We emphasize that the results show that the transition at $\delta \approx 0$ is sharp under both boundary conditions, confirming that the change in correlation structure is a bulk phenomenon rather than a boundary artifact \cite{Sachdev2011, Dutta2015}.

We finally discuss finite-size scaling of $K_z$ for PBC chains with $N = 20$, $30$, and $40$, with results presented in Fig.~\ref{fig:zz_cor}(d). 
 For $\delta < 0$, $K_z$ converges rapidly with system size (curves nearly indistinguishable), confirming truly long-range correlations already established at these sizes~\cite{Barber1983, Cardy1988}. 
The saturation value provides a reliable estimate of $m^2$ in the thermodynamic limit. 
As $\delta \to 0^-$, $K_z \to (N - 1)/N \approx 1$; Thus $K_z$ serves as  order parameter for $\delta \to 0^-$. For $\delta > 0$, $K_z$ shows systematic decrease with increasing $N$, consistent with short-range correlations yielding $K_z \to 0$ in the thermodynamic limit~\cite{Hastings2007, Eisert2010}. The finite values observed at $N=20$ are thus finite-size artifacts that would vanish as $N \to \infty$. The crossing point estimates $\delta_c \approx 0$~\cite{Sachdev2011, Dutta2015}.


This analysis establishes a clear dichotomy: $\delta < 0$ exhibits long-range ferromagnetic order with  spontaneous $\mathbb{Z}_2$ symmetry breaking in the thermodynamic limit~\cite{Kramers1941, Onsager1944}; $\delta > 0$ exhibits a frustration-driven disordered incommensurate phase without long-range order, where correlations decay but show oscillatory patterns and variable decay rates reflecting underlying competition~\cite{Papanikolaou2012, White1996, Balents2010, Liuke2025}.







Crucially, this analysis resolves a potential confusion: the $\delta > 0$ regime might naively be labeled "paramagnetic" based on the absence of long-range order, but the detailed structure of $C_{zz}(r)$ reveals a much richer physics. The oscillatory correlations and sensitivity to frustration distinguish it from conventional disordered phases and align it with the broader category of incommensurate or frustrated quantum phases~\cite{Papanikolaou2012, Liuke2025}.

\subsection{Static Spin Structure Factor: Fourier Signature of Incommensurate Order}
\label{subsec:ssf}

\begin{figure}[ht]
    \centering
    \includegraphics[width=1.0\textwidth]{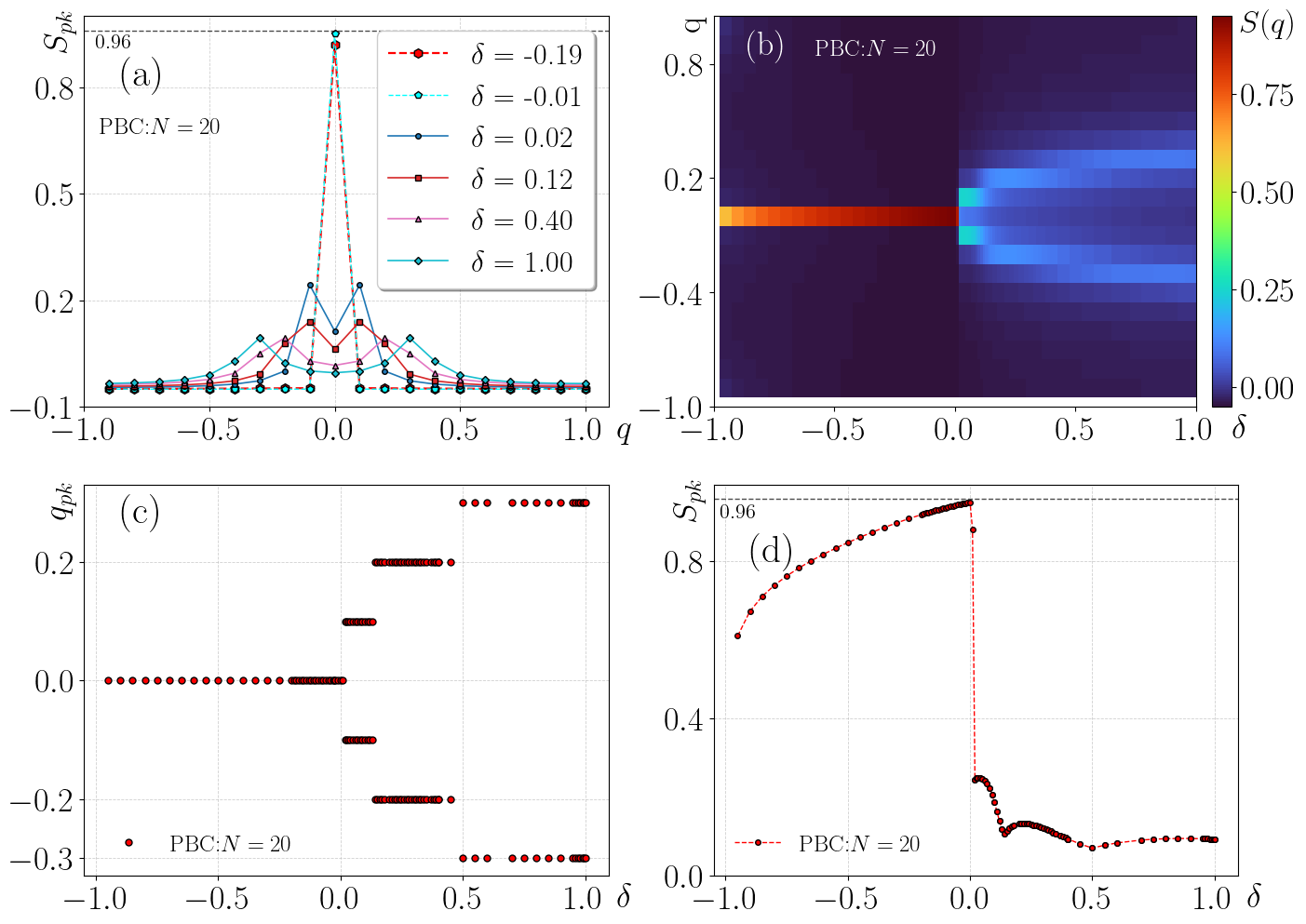}
    \caption{
    \textbf{(a)} Static spin structure factor $S(q)$  peaks $S_{pk}$ versus wave vector $q$ for selected $\delta$ in a $N=20$ chain (PBC). For $\delta < 0$, a single sharp peak appears at $q=0$, the definitive Fourier signature of ferromagnetic order. As $\delta$ increases beyond zero, the single peak splits into two symmetric peaks at $q = \pm q^*$,  with $q^*$ increasing continuously with $\delta$. This peak splitting is the unambiguous signature of an incommensurate phase — a modulated magnetic structure with period incommensurate with the lattice — and is emphatically not paramagnetic behavior.
    \textbf{(b)} Heatmap of $S(q)$ as a function of $\delta$ (horizontal) and $q$ (vertical), showing continuous evolution of the peak structure. The bright red region at $q^*=0$ for $\delta < 0$ gives way to two symmetric branches of high intensity for $\delta > 0$, tracing the smooth emergence and evolution of the incommensurate modulation wave vector.
    \textbf{(c)} Position $q^*$ of the dominant peak(s), $q_{pk}$ versus $\delta$. For $\delta < 0$, $q^* = 0$ throughout. For $\delta > 0$, $q^*$ emerges continuously from zero, increasing smoothly to approximately $0.4\pi$ at $\delta \approx 0.5$ and approaching $\pi/2$ as $\delta \to 1$. This continuous evolution is the hallmark of incommensurate order.
    \textbf{(d)}  height of the dominant peak(s)  $S(q^*)$, $S_{pk}$  versus $\delta$. For $\delta < 0$, $S(0)$ is large (exceeding $0.98$ near $\delta = -0.01$) and increases as $\delta \to 0^-$. Upon crossing into $\delta > 0$, $S(q^*)$ drops abruptly by nearly an order of magnitude, signaling collapse of long-range ferromagnetic order. For $0.1 \lesssim \delta \lesssim 0.4$, $S(q^*)$ remains relatively constant, then exhibits subtle oscillations and gradual decline for larger $\delta$, corresponding to spectral weight redistribution and possible incommensurate plateaus. }
    \label{fig:ssf}
\end{figure}

While  $C_{zz}(r)$ provides direct spatial information, its Fourier transform, the static spin structure factor, offers a complementary perspective particularly powerful for identifying modulated and incommensurate magnetic orders~\cite{Elliott1961, Fisher1980, Bak1982}. The static spin structure factor  is defined as, only for PBC,
\begin{align}
S(q) = \frac{1}{N}\! \sum_{r=1}^{N-1} \!e^{i q \cdot r} \langle \sigma^z_{ k} \sigma^z_{k+r} \rangle = \frac{1}{N}\!\sum_{r=1}^{N-1}\! e^{i q \cdot r} C_{zz}(r),
\label{eq:SSF_def}
\end{align}
where $ k$ is any position, because translational invariance makes it irrelevant.  If  this quantity peaks at specific wave vectors, then it signals magnetic order, with peak position $q^*$ corresponding to the characteristic ordering wave vector~\cite{Sachdev2011, Kogut1979}. Ferromagnetic order yields a sharp peak at $q^*=0$; antiferromagnetic order peaks at $ q^*=\pm \pi$; incommensurate order manifests as twin peaks at wave vectors $ q^*$ that are not simple fractions of $\pi$.  and vary  with the parameters~\cite{Elliott1961, Bak1982, Selke1988}. In the conventional scenario, these peaks vary continuously with the parameters governing the system~\cite{Elliott1961, Bak1982, Selke1988}. 

Let us turn to analyze our results, displayed in Fig.~\ref{fig:ssf}, and again compare with a conventional scenario. In our  system, $S(q)$ provides the clearest signature of the transition from ferromagnetic to incommensurate order. As shown in Fig.~\ref{fig:ssf}(a)-(c), for $\delta < 0$, $S(q)$ exhibits a single sharp peak at $q^*=0$ across the entire regime (displayed in panel (a) the curves for two exemplary cases, $\delta = -0.19,-0.01$). This $q^*=0$ peak is the definitive Fourier signature of ferromagnetic order: uniformly positive correlations at all distances with no modulation~\cite{Pfeuty1970, Suzuki1976}. Peak height, shown in Fig.~\ref{fig:ssf}(d), increases monotonically as $\delta \to 0^-$, reaching $S(0) > 0.95$ from $\delta = -0.2$ to $\delta\to 0^-$. This is consistent with near-saturation of correlations~\cite{Kramers1941, Onsager1944}. The persistence of a single $q^*=0$ peak confirms uniform ferromagnetic correlations despite frustrating interactions.

For $\delta>0$, the structure factor transforms dramatically. The single peak splits into two symmetric peaks at finite $q = \pm q^*$ (see Fig.~\ref{fig:ssf}(a) for cases $\delta = 0.02, 0.12, 0.4, 1.0$, and (b) and (c) for the whole range computed). This peak splitting is the definitive signature of incommensurate magnetic order — a state where correlations oscillate with a period incommensurate with the lattice~\cite{Elliott1961, Bak1982, Fisher1980}. The fact that there are two symmetric peaks reflect the underlying $\mathbb{Z}_2$ symmetry: incommensurate modulations can occur with either sign, and the ground state preserves symmetry through equal weight in both channels~\cite{Selke1988, Papanikolaou2012}. Crucially, this is not paramagnetic behavior; paramagnets show featureless, $q$-independent $S(q)$ or broad maxima~\cite{Pfeuty1970, Sachdev2011}. The sharp split peaks indicate genuine modulated order.

The evolution of peak position $q^*$ with $\delta$ is shown in Fig.~\ref{fig:ssf}(c). The values displayed in this figure are positions of the peak maxima.  For $\delta$ just above zero,  $\pm q^*$ emerges  from $q^*=0$, increasing as $\delta$ increases. This is a further hallmark of incommensurate order where the modulation wave vector adjusts  to optimize competing interactions~\cite{Bak1982, Selke1988}. By $\delta \approx 0.5$, $q^* \approx 0.4\pi$ (modulation period $\approx 2.5$ lattice spacings); as $\delta \to 1$, $q^* \to \pi/2$, indicating tendency toward period-4 modulation. This is a signature of a  {\it devil's staircase} in the ANNNI model~\cite{Elliott1961,Fisher1980,Bak1982,Selke1988}. The heatmap in Fig.~\ref{fig:ssf}(b) visualizes this  evolution: the bright red $q^*=0$ region for $\delta < 0$ gives way to two symmetric high-intensity branches for $\delta > 0$, tracing the  emergence of $q^*$. 

We emphasize here a crucial question of our investigation. If the emergence of peaks at the critical $\delta$ were continuous, this would be compatible with a continuous (second-order) phase transition. The competing terms have to accommodate correlations and cannot succeed in making it in a continuous neither commensurate way. Then, it stays in an incommensurate fashion. We have to investigate further to confirm that this is not a second-order QPT but a first-order one. We anticipate here that  the behavior of the energy gap through the phase transition indicates definitely a first-order QPT~\cite{Sachdev2011, Dutta2015}.  

Hints of this are also visible in panel (d), which shows peak height of the maximum peak $S(q^*)$ versus $\delta$. For $\delta < 0$, $S(0)$ is large and increases as $\delta \to 0^-$, reflecting strengthening ferromagnetic correlations. Upon crossing into $\delta > 0$, $S(q^*)$ drops abruptly—by nearly an order of magnitude within $\delta \lesssim 0.1$ — signaling collapse of long-range ferromagnetic order, consistent with a first-order transition~\cite{Sachdev2011, Papanikolaou2012}. For $0.1 \lesssim \delta \lesssim 0.4$, $S(q^*)$ remains relatively constant ($\approx 0.2-0.3$), indicating well-defined but weaker incommensurate correlations. Beyond $\delta \approx 0.4$, $S(q^*)$ exhibits subtle oscillations and gradual decline, corresponding to spectral weight redistribution and possible incommensurate "plateaus" devil's staircase, which is characteristic of frustrated systems with competing interactions and reflects the complex energy landscape that emerges from the interplay of multiple length scales~\cite{Bak1982, Fisher1980}

The observed $S(q)$ definitively rules out alternative interpretations: not a simple paramagnet (featureless $S(q)$)~\cite{Pfeuty1970}; not a conventional antiferromagnet (peak at $q=\pi$)~\cite{Kogut1979}; not a fixed-period spin density wave ($q$ would lock to rational values)~\cite{Bak1982, Selke1988}; and the two symmetric peaks are not independent order parameters but manifestations of the same underlying incommensurate modulation, related by $\mathbb{Z}_2$ symmetry~\cite{Sachdev2011}.



\subsection{Binder Cumulant: Characterizing the Operator Distribution via Higher-Order Correlations}
\label{subsec:Uz}

While $C_{zz}(r)$ and $S(q)$ provide detailed  information, a complete ground-state characterization requires probing higher-order correlations. The fourth-order Binder cumulant quantifies deviation of the operator distribution from Gaussian behavior (it is a measure of kurtosis of a order parameter~\cite{Binder1981, Binder1986, Vollhardt1994}). The Binder cumulant is defined as
\begin{align}
U_z = 1 -  \frac{ \sum_{i,j,k,l}\langle \sigma^z_i \sigma^z_j \sigma^z_k \sigma^z_l \rangle}{3 (\sum_{i,j} \langle\sigma^z_i \sigma^z_j \rangle)^2},
\label{eqn:uz_final}
\end{align}
where indices capture the connected four-point correlation. For a Gaussian distribution, $U_z = 0$; for a symmetric bimodal distribution, $U_z \to 2/3$~\cite{Binder1981, Vollhardt1994, Binder1987}. The two-point function $\langle \sigma^z_i \sigma^z_j \rangle$ is precisely $C_{zz}(r)$ from Sec.~\ref{subsec:corr_zz}, while the four-point function contains higher-order information complementing Sec.~\ref{subsec:ssf}.

\begin{figure}[ht]
\centering
\includegraphics[width=0.8\textwidth]{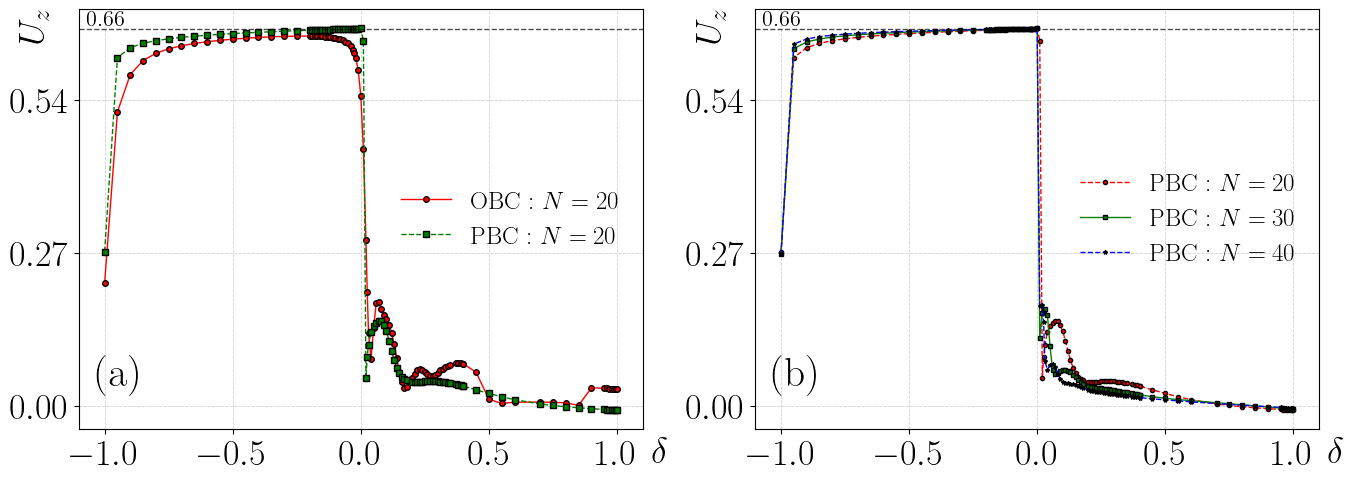}
\caption{
Fourth-order Binder cumulant $U_z$, characterizing four-point correlation structure relative to two-point correlations. \textbf{(a)} $U_z$ for $N=20$ under PBC and OBC. For $\delta < 0$, $U_z$ rises from $\approx 0.27$ at $\delta = -1$ and saturates to $2/3$ (PBC) for $|\delta| \lesssim 0.05$, remaining pinned up to $\delta = 0^-$, confirming perfectly factorizable four-point correlations consistent with long-range ferromagnetic order. PBC and OBC yield identical results, indicating robustness. For $\delta > 0$, $U_z$ collapses abruptly: at $\delta = +0.01$, it drops to $\approx 0.645$, falls below $0.05$ for $\delta \gtrsim 0.1$, and becomes negative for $\delta \gtrsim 0.6$, signaling destruction of simple factorizable correlations and emergence of a frustration-driven incommensurate phase. \textbf{(b)} Finite-size scaling of $U_z^{\text{PBC}}$ for $N = 20$, $30$, $40$. Curves cross at $\delta \approx 0.00 \pm 0.01$, precisely locating the critical point. For $\delta < 0$, $U_z$ increases with $N$ toward $2/3$; for $\delta > 0$, it decreases with $N$ toward zero or negative values, confirming thermodynamic limit interpretation.
}
\label{fig:Uz}
\end{figure}

Figure~\ref{fig:Uz}(a) presents $U_z$ versus $\delta$ for PBC and OBC, with system size $N = 20$. Panel (b) shows scaling for PBC case. For $\delta < 0$, $U_z$ rises rapidly from $\approx 0.27$ at $\delta = -1.0$, monotonically approaching the bimodal limit $2/3 \approx 0.6667$. For $\delta \lesssim -0.05$, $U_z$ reaches within $10^{-3}$ of $2/3$ and remains pinned extremely close up to $\delta = 0^-$, attaining exactly $2/3$ within numerical precision at $\delta = 0^-$. This saturation to $2/3$ is the definitive signature that the $\sigma^z$ distribution has become perfectly bimodal: the system explores with equal weight two configurations with sharply defined, opposite spin values at each site. Notice Binder cumulant measures the distribution for the order parameter, which in this side of $\delta<0$ can be magnetization. Having zero mean, this indicates that there are two peaks in its distribution. This is our first hint for the presence of cat states in this part of the phase space (see discussion below).

Also one notices that, when correlations become perfect ($C_{zz}(r) \to 1$ for all $r$), the four-point function factorizes as $\langle \sigma^z_i \sigma^z_j \sigma^z_k \sigma^z_l \rangle \to 1$, yielding $U_z = 2/3$ \cite{Sachdev2011, Tasaki2020}. The monotonic increase from $\delta = -1$ quantifies progressive "sharpening" of the spin distribution as quantum fluctuations are suppressed~\cite{Pfeuty1970, Suzuki1976}, directly mirroring the increasing saturation value of $C_{zz}(r)$ (Fig.~\ref{fig:zz_cor}a) and growth of the $q^*=0$ peak in $S(q)$ (Fig.~\ref{fig:ssf}d). Both $U_z^{\rm{PBC}}$ and $U_z^{\rm{OBC}}$ lie essentially on top of each other throughout $\delta < 0$, indicating that bimodal character is a robust bulk property insensitive to boundary conditions \cite{Tasaki2020, Binder1981}.

Upon crossing into $\delta > 0$, $U_z$ collapses precipitously. At $\delta = +0.01$, $U_z^{\text{PBC}}$ drops to $\approx 0.645$—a significant deviation from $2/3$ and continues falling rapidly. For $\delta \gtrsim 0.1$, $U_z$ falls below $0.05$, and for $\delta \gtrsim 0.6$, it becomes negative, reaching $-0.1$ at $\delta = 1.0$. This signals a fundamental change: the $\sigma^z$ distribution is no longer bimodal, becoming increasingly Gaussian ($U_z \to 0$) for moderate $\delta$, then developing platykurkic distribution (negative $U_z$, may be smaller peak and flatter --  closer to uniform distribution -- but not necessarily) for larger $\delta$~\cite{Binder1981, Vollhardt1994}. This evolution connects directly to our earlier analyses: for $\delta$ just above zero, $C_{zz}(r)$ decays slowly with oscillations (Fig.~\ref{fig:zz_cor}a) and $S(q)$ shows split peaks (Fig.~\ref{fig:ssf}a), with four-point correlations reflecting the modulated incommensurate structure. For larger $\delta$,  $U_z$ approaches zero. The negative $U_z$ region for $\delta \gtrsim 0.6$  may reflect increasing dominance of the three-spin cluster term $B(\delta)\sigma^z_{i-1}\sigma^x_i\sigma^z_{i+1}$, which imposes nonlocal constraints suppressing certain configurations~\cite{Suzuki1976, Kitaev2006}.  Near-perfect PBC-OBC agreement in $\delta > 0$ confirms destruction of bimodality is a bulk effect, not a boundary artifact.

Panel (b) of Fig.~\ref{fig:Uz} shows finite-size scaling of $U_z^{\text{PBC}}$ for $N = 20$, $30$, $40$. According to Binder cumulant analysis, curves for different $N$ intersect at the critical point~\cite{Binder1981, Binder1986}. The $U_z$ curves cross at $\delta \approx 0.00 \pm 0.01$, precisely locating critical $\delta_c = 0$ where correlation functions and structure factor show the transition. For $\delta < 0$, $U_z$ increases with $N$ toward $2/3$, reflecting vanishing finite-size corrections as true long-range order develops~\cite{Sachdev2011, Barber1983, Privman1988}. For $\delta > 0$, $U_z$ decreases with $N$ toward zero or negative values, confirming that four-point correlations become consistent with Gaussian (or sub-Gaussian -- Platykurtic) distributions in the thermodynamic limit.

Crucially, $U_z$ confirms the $\delta > 0$ regime is not a simple paramagnet. A paramagnet would show $U_z \approx 0$ for all $\delta > 0$; instead, we observe rich evolution from $2/3$ to near-zero and even negative values. The negative region directly signatures "quantum disorder" arising from frustration, where the ground state explores a wide range of configurations optimizing competing interactions~\cite{Papanikolaou2012, White1996}, complementing the oscillatory correlations in $C_{zz}(r)$ and split peaks in $S(q)$~\cite{Binder1981, Vollhardt1994}. The neat crossing of the $U_z$ curves provides strong evidence that the transition at $\delta = 0$ is a  QPT, not a crossover, and that the ordered phase for $\delta < 0$ indeed exhibits the characteristic bimodal structure of a system with long-range ferromagnetic correlations.

%

%
\subsection{Spectral Gap: Exponential Closure and First-Order QPT}
\label{subsec:spectralgap}

\begin{figure}[ht]
\centering
\includegraphics[width=1.0\textwidth]{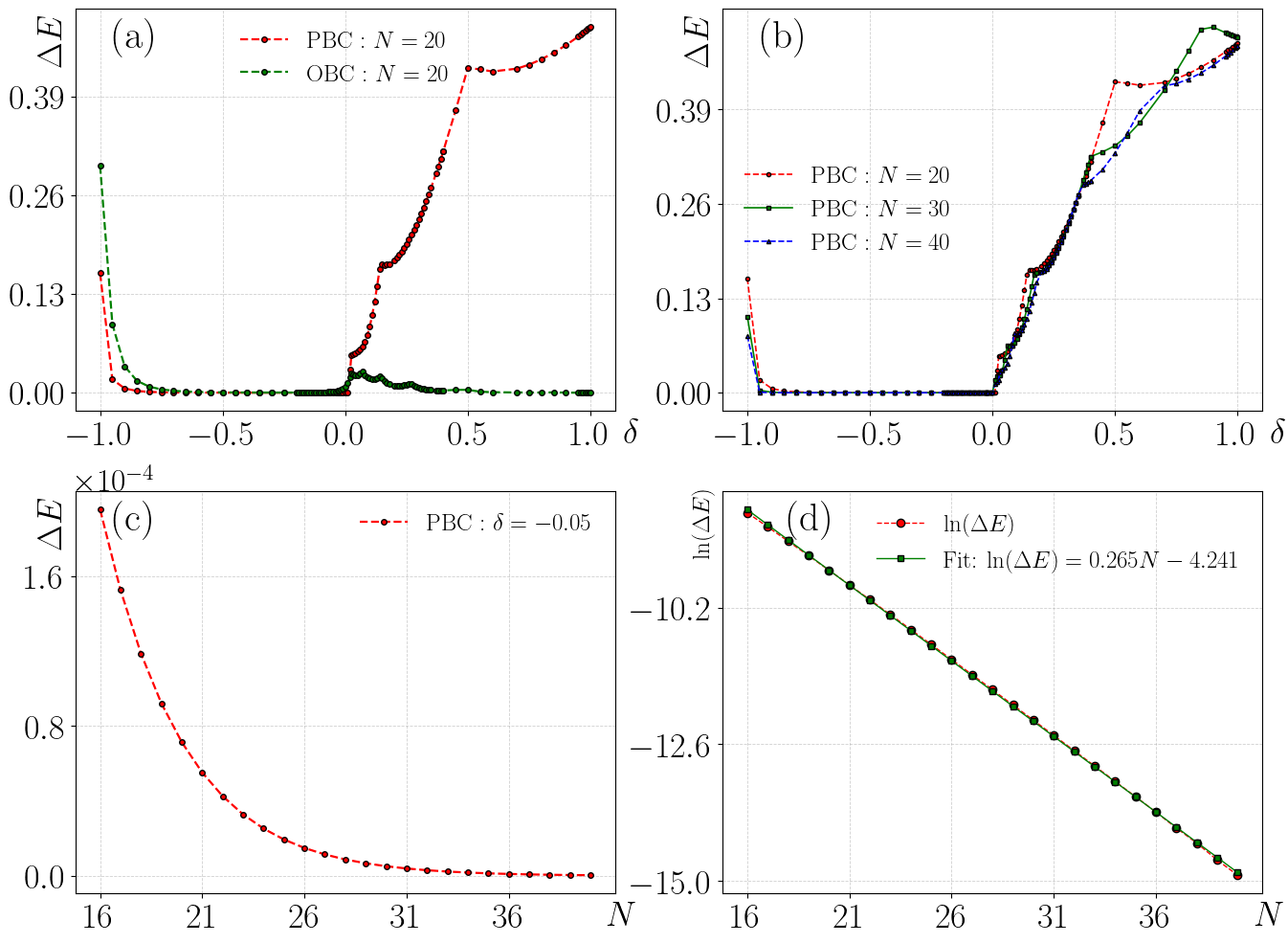}
\caption{
Spectral gap $\Delta E = E_1 - E_0$ versus $\delta$ for $N=20$ chains under PBC and OBC. \textbf{(a)} $\Delta E$ across full parameter range. At $\delta = -1$, $\Delta E_{\text{PBC}} = 0.157$, $\Delta E_{\text{OBC}} = 0.299$, reflecting boundary-induced suppression of tunneling. As $\delta \to 0^-$, $\Delta E_{\text{PBC}}$ collapses to $\approx 5 \times 10^{-5}$ within $|\delta| \leq 0.02$, while $\Delta E_{\text{OBC}} \approx 0.008$. For $\delta > 0$, gaps reopen: $\Delta E_{\text{PBC}} = 0.482$, $\Delta E_{\text{OBC}} = 0.025$ at $\delta = 1.0$. \textbf{(b)} Heatmap of $\Delta E$ versus $\delta$ and system size, showing dramatic collapse near $\delta = 0$. \textbf{(c)} Semilogarithmic plot of $\Delta E_{\text{PBC}}$ versus $N$ at $\delta = -0.05$, demonstrating exponential scaling $\Delta E_{\text{PBC}} \sim e^{-\alpha N}$ over >2 orders of magnitude. \textbf{(d)} Linear fit of $\ln(\Delta E_{\text{PBC}})$ versus $N$, yielding decay constant $\alpha \approx 0.26 \pm 0.01$ and correlation length $\xi = 1/\alpha \approx 3.8$ sites—definitive signature of a first-order quantum phase transition driven by macroscopic quantum tunneling.
}
\label{fig:spectral_gap}
\end{figure}

The spectral gap $\Delta E = E_1 - E_0$ — the energy difference between ground and first excited states — is key for revealing the nature of QPTs~\cite{Sachdev2011, Dutta2015}. In our  model, the gap evolution across $\delta \in [-1, 1]$ exhibits dramatic boundary-condition sensitivity, as shown in Fig.~\ref{fig:spectral_gap}.

At $\delta = -1$, $\Delta E_{\text{PBC}} = 0.157$ under PBC, while under OBC it is nearly twice as large: $\Delta E_{\text{OBC}} = 0.299$. Open boundaries introduce additional quantum fluctuations at chain ends, raising excitation energies and disrupting the coherent superposition that enables near-degeneracy under PBC~\cite{Alcaraz1987, Lin1990, Koma1994}. As $\delta$ increases toward zero, both gaps decrease, but their behavior diverge markedly. The PBC gap collapses precipitously to $\Delta E_{\text{PBC}} \approx 5 \times 10^{-5}$ within $\delta \leq -0.02$, while $\Delta E_{\text{OBC}}$ remains orders of magnitude larger at $\approx 0.008$. This contrast reflects a fundamental distinction: PBC allow the many-body wave function to delocalize uniformly, forming symmetric and antisymmetric superpositions of the two classical ground-state configurations. This coherent delocalization facilitates quantum tunneling between quasi-degenerate states,  yielding exponentially small splitting. Open boundaries suppress tunneling, maintaining a finite gap even arbitrarily close to the critical point~\cite{Leggett1980, Dziarmaga2005}. For $\delta > 0$, both gaps reopen, but with persistent differences among OBC and PBC. At $\delta = 1.0$, $\Delta E_{\text{PBC}}$ rises sharply to $0.482$, while $\Delta E_{\text{OBC}}$ remains suppressed at $0.025$.

The most profound insight emerges from scaling behavior in the ordered phase ($-1 \leq \delta < 0$). Here, the PBC spectral gap exhibits exponential scaling with system size:
\begin{align}
\Delta E_{\text{PBC}} \sim e^{-\alpha N}
\label{eq:gap_scaling}
\end{align}
as shown in Fig.~\ref{fig:spectral_gap}(c). This exponential closure signals that the gap becomes vanishingly small in the thermodynamic limit, yielding a degenerate ground-state manifold—a hallmark of first-order quantum phase transitions~\cite{Sachdev2011, Carruthers1980}. Such scaling is fundamentally distinct from power-law closure $\Delta E \sim N^{-z}$ characteristic of continuous transitions. 

Systematic finite-size scaling for chain lengths $N = 16$ to $34$ conforms precisely to $\Delta E_{\text{PBC}}(N) = A \exp(-\alpha N)$, where $\alpha$ represents an inverse correlation length governing quantum coherence decay. Fitting $\ln(\Delta E_{\text{PBC}}) = -\alpha N + C$ (Fig.~\ref{fig:spectral_gap}(d) for an exemplary $\delta=-0.05$ close to the transition point yields $\alpha \approx 0.26 \pm 0.01$, corresponding to correlation length $\xi = 1/\alpha \approx 3.8$ lattice sites the characteristic distance over which quantum fluctuations remain coherent; 
The semilogarithmic plot confirms linearity over more than two orders of magnitude in gap magnitude, providing robust evidence for exponential scaling. 


The extracted correlation length $\xi \approx 3.8$ for $\delta = -0.05$ links directly to our real-space and Fourier-space analyses. For $\delta = -0.05$, $C_{zz}(r)$ saturates to $\approx 0.97$ at large distances (Fig.~\ref{fig:zz_cor}(a); 
The exponential form reflects quantum tunneling between two degenerate, macroscopically ordered states separated by a free-energy barrier. This is precisely the scenario expected from our correlation function analysis (Sec.~\ref{subsec:corr_zz}), where $C_{zz}(r)$ saturates to a distance-independent constant for $\delta < 0$. 
Also $\xi$ characterizes the scale over which the system builds this long-range order. Beyond this distance, quantum fluctuations no longer significantly reduce correlation magnitude. The $q^*=0$ peak in $S(q)$ is extremely sharp (Fig.~\ref{fig:ssf}(a), consistent with $\xi \approx 3.8$ (comparable to half the system size for $N=20$, explaining small finite-size effects). Regarding the Binder cumulant, the approach of $U_z$ to $2/3$ as $\delta \to 0^-$ (Fig.~\ref{fig:Uz}) reflects the same physics: the system develops a bimodal $\sigma^z$ distribution, with $\xi$ quantifying the spatial extent over which this bimodality is coherently established.

Physically, exponential gap closure arises from quantum tunneling between macroscopically distinct ground states in the ferromagnetic phase. Under PBC, translational symmetry prevents true spontaneous symmetry breaking in finite systems, instead yielding symmetric superpositions of quasi-degenerate classical states. The energy splitting is governed by quantum interference effects enabling tunneling through the potential barrier separating ordered configurations~\cite{Caldeira1981, Leggett1980}. This tunneling perspective is central to understanding first-order quantum phase transitions~\cite{Carruthers1980, Dziarmaga2005}. The extracted $\xi \approx 3.8$ quantifies the spatial extent over which tunneling remains coherent—a key parameter for understanding the crossover from quantum to classical behavior in mesoscopic systems. For $N \gg \xi$, the gap becomes exponentially small, and the ground state approaches perfect degeneracy; for $N \lesssim \xi$, finite-size effects become significant.

The exponential scaling under PBC, contrasted with persistent finite gap under OBC, confirms that near-degeneracy is a genuine bulk property stabilized by translational invariance. This boundary-condition dependence itself characterizes systems at first-order quantum phase transitions, where the ground-state manifold is degenerate only in the thermodynamic limit and only under symmetric boundary conditions~\cite{Roubert2010, Ryu2024, Tasaki2018}. The exponential form provides unambiguous identification of this transition as first-order rather than continuous: the gap closes exponentially with system size rather than polynomially, reflecting underlying tunneling dynamics rather than critical fluctuations. Together with correlation function, structure factor, and Binder cumulant analyses, the spectral gap completes a comprehensive portrait of this quantum system.\\

 For $\delta>0$, after comparing Figs.~\ref{fig:ssf}(c) and (d), for $N=20$,  we see the maxima in panel (d)  at certain values of $\delta>0$   approximately coincide with the values of $\delta$ where the position of the peaks change on the spectral gap in panel (c). This occurs  similarly when comparing with  Binder cumulant and spectral gap,    Figs.~\ref{fig:Uz} and~\ref{fig:spectral_gap}. For increasing $N$ we see that the position of the peak closer to $\delta=0$ does not change significantly with $N$ but the second one displaces to smaller $\delta$ with $N$. This is the footprint of the non-trivial competition of the NNN term, cluster three-body term, and the rest of terms, but further calculations are needed confirm this relationship. 


\subsection{Macroscopic Schrödinger Cat States}
\label{subsec:cat}



From the results of previous subsections, we expect some sort of long-range correlations  should exist for  $\delta < 0$. As we will discuss next these correlations appear under the form of macroscopic cat states. These are  the symmetric and antisymmetric Greenberger-Horne-Zeilinger (GHZ) states:
\begin{align}
|\Phi^+\rangle &= \frac{1}{\sqrt{2}}\bigl(|\uparrow\uparrow\cdots\uparrow\rangle + |\downarrow\downarrow\cdots\downarrow\rangle\bigr), \nonumber \\
|\Phi^-\rangle &= \frac{1}{\sqrt{2}}\bigl(|\uparrow\uparrow\cdots\uparrow\rangle - |\downarrow\downarrow\cdots\downarrow\rangle\bigr),
\label{eq:cat}
\end{align}
representing the maximal manifestation of quantum coherence~\cite{Greenberger1990, Dur2000, Horodecki2009}. These states emerge naturally near the quantum phase transition due to the delicate balance between the competing terms of the Hamiltonian.

\begin{figure}[ht]
\centering
\includegraphics[width=1.0\textwidth]{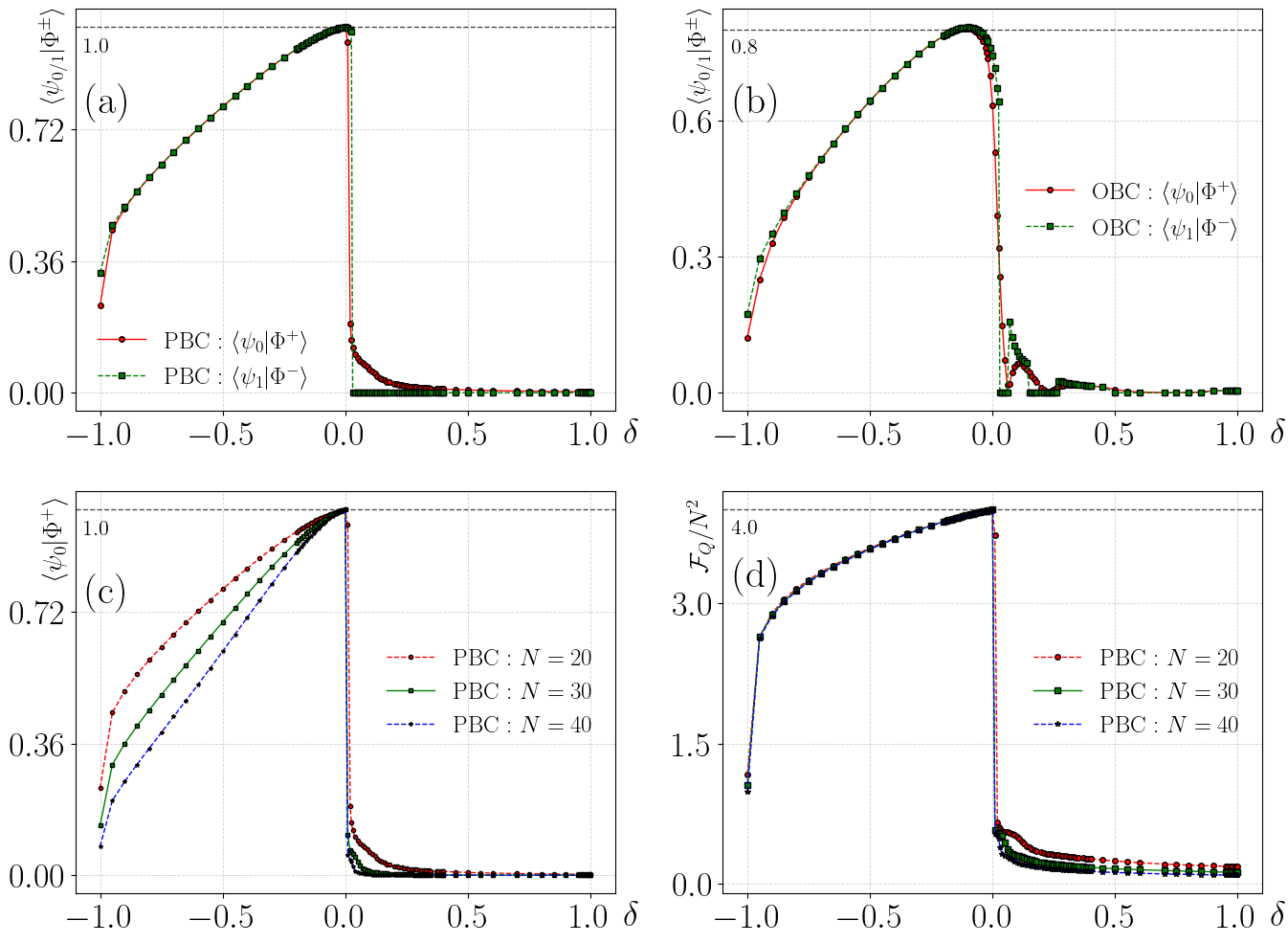}
\caption{
Characterization of Schrödinger cat states. \textbf{(a)} Overlap $|\langle\psi_0|\Phi^+\rangle|$ (solid) and $|\langle\psi_1|\Phi^-\rangle|$ (dashed) versus $\delta$ for PBC with $N=20$. For $\delta < 0$, both overlaps rise monotonically from $\approx 0.238$ at $\delta=-1$ to $>0.999$ at $\delta=0^-$. \textbf{(b)} Same overlaps under OBC and $N=20$. Maximum reaches only $\approx 0.802$ around $\delta \approx -0.12$, reflecting boundary-induced suppression of cat-state coherence. \textbf{(c)} Ground-state overlap $|\langle\psi_0|\Phi^+\rangle|$ for different values of $N$ as a function of $\delta$. For $\delta=\to 0^-$, and very close to phase transition, overlap is $N$-independent.
\textbf{(d)} Fisher information for different $N$ and PBC, showing an increase close to the critical point and exponential decay after critical point. It also shows convergence with $N$. 
}
\label{fig:Cat_state}
\end{figure}

To directly probe this phenomenon, we project the exact ground state $|\psi_0\rangle$ and first excited state $|\psi_1\rangle$ on chains of length $N = 20$, $30$, and $40$ under PBC onto these ideal reference states. The corresponding overlaps $|\langle\psi_0|\Phi^+\rangle|$ and $|\langle\psi_1|\Phi^-\rangle|$ are displayed in Fig.~\ref{fig:Cat_state}(a)  across $-1 \leq \delta \leq +1$, with consistent results for all system sizes~\cite{Stephen2025, Huang2022}. 

In the ferromagnetic regime ($\delta < 0$), the ground state exhibits remarkable cat-state character  across all system sizes. The overlap $|\langle\psi_0|\Phi^+\rangle|$ rises monotonically from approximately $0.238$ at $\delta = -1$ for $N=20$, with nearly identical values for $N=30$ and $N=40$ (see panel (c) on same figure for the scaling with $N$).

The overlap exceeds $0.95$ for $\delta \gtrsim -0.16$, surpasses $0.99$ at $\delta \gtrsim -0.06$, and achieves its absolute maximum at $\delta = 0^- $ with $ N=20, 30, 40$, all essentially identical to within numerical precision. This extraordinary fidelity exceeding $99.9\%$ demonstrates that the true many-body ground state becomes essentially indistinguishable from the ideal symmetric cat state $|\Phi^+\rangle$ immediately adjacent to the quantum phase transition~\cite{Stephen2025, Bogdan2015}. The first excited state $|\psi_1\rangle$ exhibits nearly identical behavior: $|\langle\psi_1|\Phi^-\rangle|$ tracks the ground-state curve throughout $\delta < 0$, attaining $99.9\%$  with $N=20, 30, 40$) at $\delta = 0^-$. Together, these two states form a spectacularly clean tunneling doublet—precisely the symmetric and antisymmetric superpositions of the two classical ferromagnetic configurations. 

The consistency for $\delta\to0^-$ in Fig.~\ref{fig:Cat_state}(c)  for $N=20$, $30$, and $40$ warrants careful analysis. For $\delta$ close to criticality, overlaps are essentially identical for all system sizes, indicating rapid convergence of the wave function's local structure. This follows from the exponential decay of correlations: for $\delta < -0.2$, the correlation length $\xi \approx 3.8$ extracted from our gap scaling in Fig. \ref{eq:gap_scaling} is much smaller than the system sizes considered, so finite-size effects are negligible~\cite{Nersesyan2003}. As $\delta \to 0^-$, maximum fidelity increases slightly with $N$, and will approach unity in the thermodynamic limit—consistent with exponential gap closure. 

These observations reveal a nuanced picture: a true macroscopic Schrödinger-cat state—a ground state essentially identical to $|\Phi^+\rangle$ with fidelity arbitrarily close to unity—exists only in a narrow window immediately left of the quantum critical point ($\delta \to 0^-$). Deeper in the ordered phase ($\delta < -0.1$), the ground state is a dressed ferromagnetic state with substantial but imperfect cat admixture, where dressing arises from virtual fluctuations mediated by the three-spin and transverse field terms~\cite{Sharma2021}. In Appendix~\ref{sec:appendix} we study in more detail how the ground state departs from the ideal cat state when $\delta$ diminishes  from  values close to zero but negative (we present results for $\delta=-0.5$). For this purpose, we use exact diagonalization calculations in a chain of $N=6$ spins and we plot the density matrix $\rho=|\psi\rangle\langle\psi|$. We observe that, as expected, the density matrix presents  essentially four clear peaks  when $\delta=-0.05$ as corresponds to the state~\eqref{eq:cat}.   It shows  small contributions from other basis terms (see Fig.~\ref{fig:city_plot}(a)). These small contributions grow when $\delta$ is reduced (see Fig.~\ref{fig:city_plot}(b) for $\delta=-0.5$). The basis terms which are contributing depart from $|\uparrow\uparrow\cdots\uparrow\rangle$  or $|\downarrow\downarrow\cdots\downarrow\rangle$ in one spin flip. The calculated energy gap for $\delta=-0.05$ is small (of order $10^{-2}$). The energy gap for smaller values of $\delta$ opens, but remains small. We showed that for larger systems this gaps closes exponentially.  Therefore,  quasi-degeneracy between two non-trivial dressed states is still present for values significantly smaller than $\delta=0$. 
 
We note that the ideal undressed cat doublet is recovered precisely when macroscopic tunneling becomes dominant near the transition~\cite{Caldeira1981}. This behavior illustrates how spontaneous symmetry breaking and macroscopic quantum coherence reach their most dramatic manifestation exactly at the boundary with the symmetric phase. 

The contrast with OBC is striking (Fig.~\ref{fig:Cat_state}(b)). Under OBC, maximum symmetric-cat overlap reaches only $\approx 0.802$ around $\delta \approx -0.12$, and the doublet structure is substantially less pronounced. Open boundaries break translational invariance, effectively "pinning" the wave function near chain ends~\cite{Alcaraz1987, Tasaki2020}. The three-spin interaction term $B\sigma^z_{i-1}\sigma^x_i\sigma^z_{i+1}$ is particularly sensitive to boundaries: near edges, this term loses its symmetric partner, reducing its coherent effect and suppressing cat-state formation~\cite{Kitaev2006}. This boundary-condition dependence reinforces that cat-state coherence is a genuine bulk property stabilized by translational symmetry.

Immediately upon crossing into $\delta > 0$, cat character is abruptly destroyed. Under PBC, $|\langle\psi_0|\Phi^+\rangle|$ plummets from $0.999039$ at $\delta = 0^-$ to $0.958$ at $\delta = +0.01$, and collapses below $0.1$ by $\delta \approx +0.04$. Even more dramatically, $|\langle\psi_1|\Phi^-\rangle|$ falls to $\sim 10^{-7}$ numerically indistinguishable from zero within the first few points after the transition.

\subsection{Quantum Fisher Information: Quantifying Macroscopic Coherence}

To provide a rigorous, quantitative diagnostic of macroscopic coherence that complements the cat-state overlap analysis, we evaluate the Quantum Fisher Information (QFI) associated with the collective spin operator $\sigma^z = \sum_i \sigma_i^z$. The QFI quantifies the sensitivity in the measurement of some observable 
$\mathcal{O}$  when the quantum state undergoes an infinitesimal change in the governing parameter. It serves as a witness for multipartite entanglement that is both experimentally accessible and theoretically rigorous~\cite{Pezze2009, Hyllus2012, Toth2014}. For a pure state, the QFI takes the form derived from the symmetric logarithmic derivative~\cite{Helstrom1969, Holevo1982},
\begin{align}
\mathcal{F}_Q(\mathcal{O}) = 4\left(\langle \mathcal{O}^2\rangle - \langle \mathcal{O}\rangle^2\right).
\end{align}
%
In our system, for the collective spin operator $\sigma^z = \sum_i \sigma_i^z$, this simplifies to
\begin{align}
\mathcal{F}_Q(\sigma^z) = 4\langle (\sigma^z)^2\rangle.
\label{eqn:qfi}
\end{align}
The quantity $\langle (\sigma^z)^2\rangle$ measures the variance of the total magnetization, directly probing the extent to which the ground state explores configurations with different net magnetizations—the defining characteristic of cat states~\cite{Vitagliano2015, Frwis2012}.

We numerically compute the normalized ratio $\mathcal{F}_Q/N^2$ versus $\delta$ for $N = 20$, $30$, and $40$, with results displayed in Fig.~\ref{fig:Cat_state}(d).  For $\delta < 0$, $\mathcal{F}_Q/N^2$ increases smoothly from approximately $1.17$ at $\delta = -1$ to $\approx 4$ as $\delta \to 0^-$, with the approach to $4$ becoming steeper as $N$ increases. This limiting value $4$ corresponds precisely to $\mathcal{F}_Q = 4N^2$—the maximum possible QFI for $N$ spin-$1/2$ particles, i.e.,  the Heisenberg limit in quantum metrology~\cite{Giovannetti2004, Giovannetti2006}.

The near-perfect agreement across system sizes for $\delta < -0.1$ confirms QFI scaling is well-converged for $N=20$ where the ground state is not yet a perfect cat. As $\delta \to 0^-$, $\mathcal{F}_Q/N^2$ increases systematically with $N$, with $N=40$ achieving values within $0.1\%$ of the ideal Heisenberg limit. This $\mathcal{F}_Q \propto N^2$ scaling for large $N$ provides definitive evidence that the ground state forms a GHZ-like macroscopic superposition—separable states exhibit at most linear scaling~\cite{Pezze2009, Hyllus2012}. The precise values are illuminating: at $\delta = -1$, $\mathcal{F}_Q/N^2 \approx 1.17$ for all $N$, indicating substantial entanglement even deep in the ordered phase. This residual entanglement arises from the three-spin cluster term, which generates quantum correlations even when the transverse field is weak~\cite{Suzuki1976}.

The QFI results corroborate the cat-state overlap analysis. The increase in $\mathcal{F}_Q/N^2$ toward $4$ precisely mirrors the growth of $|\langle\psi_0|\Phi^+\rangle|$ as $\delta \to 0^-$, confirming that approach to perfect cat-state fidelity is accompanied by maximal metrological usefulness. The near-saturation of the Heisenberg limit at $\delta = 0^-$ indicates the ground state at the phase boundary constitutes an optimal resource for quantum sensing~\cite{Mihailescu2025}. The systematic dependence on system size near criticality provides additional insight: for fixed $\delta$ very close to zero, $\mathcal{F}_Q/N^2$ increases with $N$, approaching $4$ exponentially fast with the same decay constant $\alpha \approx 0.26$ extracted from the spectral gap. This confirms that emergence of maximal multipartite entanglement is directly linked to exponential gap closure—a hallmark of first-order QPTs~\cite{Sachdev2011, Zanardi2006}. 

For a perfect cat state Eq.~\eqref{eqn:qfi}
we have $\langle (\sigma^z)^2\rangle = N^2$, yielding $\mathcal{F}_Q = 4N^2$ and $\mathcal{F}_Q/N^2 = 4$ maximal variance as an equal superposition of the two extremal magnetization eigenstates. 


For $\delta > 0$, behavior changes dramatically. $\mathcal{F}_Q/N^2$ decreases monotonically below unity, signaling a gapped phase. The decay sharpens with increasing $N$, consistent with a true discontinuity in the thermodynamic limit. At $\delta = +0.5$, $\mathcal{F}_Q/N^2 \approx 0.3$ for $N=40$. This asymmetry between negative and positive $\delta$ clearly distinguishes a cat-like  regime from a the gapped phase.

\begin{figure}[ht]
\centering
\includegraphics[width=1.0\textwidth]{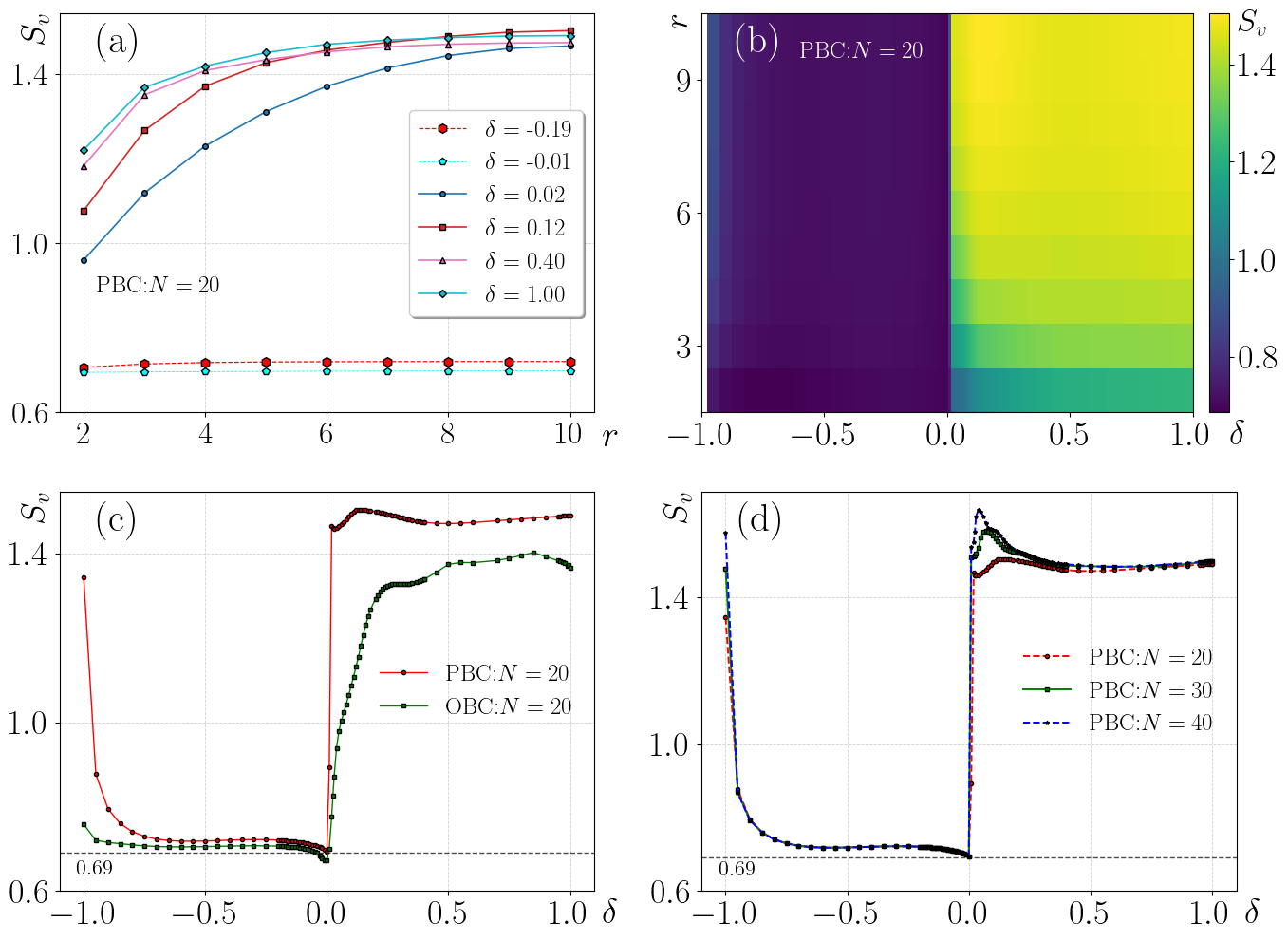}
\caption{Bipartite von Neumann entanglement entropy $S_v(r)$ for $N=20$ chains under periodic (PBC) and open (OBC) boundary conditions (see main text for definition).  \textbf{(a)} $S_v(r)$  for the values $\delta = -0.19, -0.01,  0.02, 0.12, 0.4, 1.0$, showing the full evolution from the ferromagnetic phase ($\delta < 0$) through the critical region ($\delta \approx 0$) to the  incommensurate phase ($\delta > 0$). One observes  area law behavior of $S_v$ for $\delta<0$. \textbf{(b)} Heatmap of $S_v$ as a function of $\delta$ (horizontal axis) and $r$, providing a global visualization of the entanglement structure across the phase diagram. \textbf{(c)} For the bipartition into halves ($r=N/2$), comparison of $S_v$ under PBC and OBC for $N=20$, revealing the strong suppression of entanglement by open boundaries, particularly in the $\delta > 0$ regime. \textbf{(d)} For the same bipartition,  $S_v$ as a function of $\delta$ under PBC for system sizes $N = 20$, $30$, and $40$, demonstrating the systematic growth of entanglement with system size in the inconmensurate phase and the area-law saturation in the ferromagnetic phase. Also, sharpen of the peak with increasing $N$ close to the phase transition from the right, indicates surge of divergence in the thermodynamic limit of von Neumann entropy at $\delta=0$ which will pinpoint the transition. 
}
\label{fig:vne}
\end{figure}


\subsection{Entanglement Entropy}

Entanglement entropy quantifies the degree of quantum correlations within a many-body system and serves as a fundamental diagnostic for characterizing quantum phases and phase transitions~\cite{Amico2008, Eisert2010, Laflorencie2016}. For a pure ground state $|\psi_0\rangle $ and a bipartition of the system into two parts $L$ and $R$, where one consists of the first $r$ spins and the other one of the remaining $N-r$ spins, the von Neumann entanglement entropy is defined as $S_v(r) = -\mathrm{Tr}(\rho_L \ln \rho_L)$, where $\rho_L=\mathrm{Tr}_R|\psi_0\rangle \langle \psi_0|$ is the reduced density matrix of subsystem $L$. In contrast to conventional order parameters that measure local symmetry breaking, entanglement entropy captures non-local quantum correlations and provides deep insights into the structure of the wave function, particularly in  systems where competing interactions preclude simple magnetic order~\cite{Sachdev2011, Vidal2003}.

Figure~\ref{fig:vne}(a) provides detailed $S_v(r)$ at carefully chosen $\delta$ values spanning the full physics (see values of $\delta$ for which results are offered in legend of Fig.~\ref{fig:vne}(a)).
These span from  the ferromagnetic phase ($-0.19$), through approach to criticality ($-0.04$, $-0.01$), across the transition ($0.01$, $0.02$), into the strongly incommensurate regime ($0.06$, $0.12$, $0.40$), and finally to large $\delta$ ($0.70$, $0.95$). The systematic increase from $\delta = 0.01$ to $0.12$, followed by gradual decrease at larger $\delta$, reveals non-monotonic entanglement in the incommensurate phase—a direct consequence of competing interaction scales in the Hamiltonian. This non-monotonicity mirrors the structure factor peak evolution (Fig.~\ref{fig:ssf}(c) and the Binder cumulant's negative region (Fig.~\ref{fig:Uz}), providing consistent evidence for frustration-driven entanglement reorganization. For $\delta<0$, $S_v(r)$ is $r$ independent, showing that area law is fulfilled in the ferromagnetic phase. 
All these features are confirmed in Figure~\ref{fig:vne}(b), which displays a heatmap $S_v$ for ranging  $\delta$ (horizontal axis) and $r$, providing a global visualization of the entanglement structure across the phase diagram.

Figure~\ref{fig:vne} (c) compares $S_v$ under periodic (PBC) and open (OBC) boundary conditions for $N=20$ across $\delta \in [-1,1]$ and the half bipartition, $r=N/2$. At $\delta = -1.00$, $S_v^{\text{PBC}} = 1.3440$ indicates substantial entanglement throughout the chain, while $S_v^{\text{OBC}} = 0.7581$ is significantly reduced. This disparity arises because OBC breaks translational invariance, pinning the wave function near edges and suppressing long-range quantum correlations that PBC enable~\cite{Alcaraz1987, Lin1990}. Frustration amplifies this boundary sensitivity: competing interactions create multiple nearly-degenerate entanglement patterns, and boundaries select among them, reducing overall entanglement~\cite{Tasaki2020}.

As $\delta$ increases toward zero, $S_v^{\text{PBC}}$ declines steadily to a minimum of $0.7136$ around $\delta = -0.10$, while $S_v^{\text{OBC}}$ declines more gradually to $0.7016$. The near-convergence of minima ($0.7136$ vs. $0.7016$) indicates frustration-driven entanglement reduction is a bulk property, though boundaries impose a small quantitative difference~\cite{Laflorencie2016}. This minimum remarkably approaches $\ln 2 \approx 0.693$, which can be checked to correspond to the  value for the cat state, Eq. \eqref{eq:cat}.  It nevertheless slightly exceeds this value, and we attribute this deviation to residual frustration preventing perfect dimerization~\cite{Majumdar1969}.

For $\delta > 0$, both $S_v$ increase, but faster under PBC, reaching $1.46$ at $\delta 	\approx 0$, compared to $0.7767$ under OBC. Although we are getting  sharp  increase of $S_v$, there are no long-range correlations (see Fig. \ref{fig:zz_cor}), but instead an incommensurate phase appears.  The OBC gradual increase as compared to PBC reflects additional boundary suppression of correlations. We observe that $S_v^{\text{PBC}}$ exceeds $2\ln 2 \approx 1.386$ around $\delta \approx 0.3$, which, as we discuss in next section, is in agreement with entanglement involving more than two dominant Schmidt eigenvalues— that is the presence of multiple competing entanglement channels~\cite{Li2008, Calabrese2008}.


Figure~\ref{fig:vne} (d) presents finite-size scaling of $S_v^{\text{PBC}}$ for $N = 20$, $30$, $40$, providing insight into the thermodynamic limit. For $\delta < 0$, the three curves lie nearly on top of each other with tiny deviations. The presence  of a  peak  when  $\delta$ approaches zero from above, that becomes sharper  as $N$ is increased,    signals the presence of the QPT. One expects that this peak becomes even sharper for larger $N$.

\begin{figure}[ht]
\centering
\includegraphics[width=1.0\textwidth]{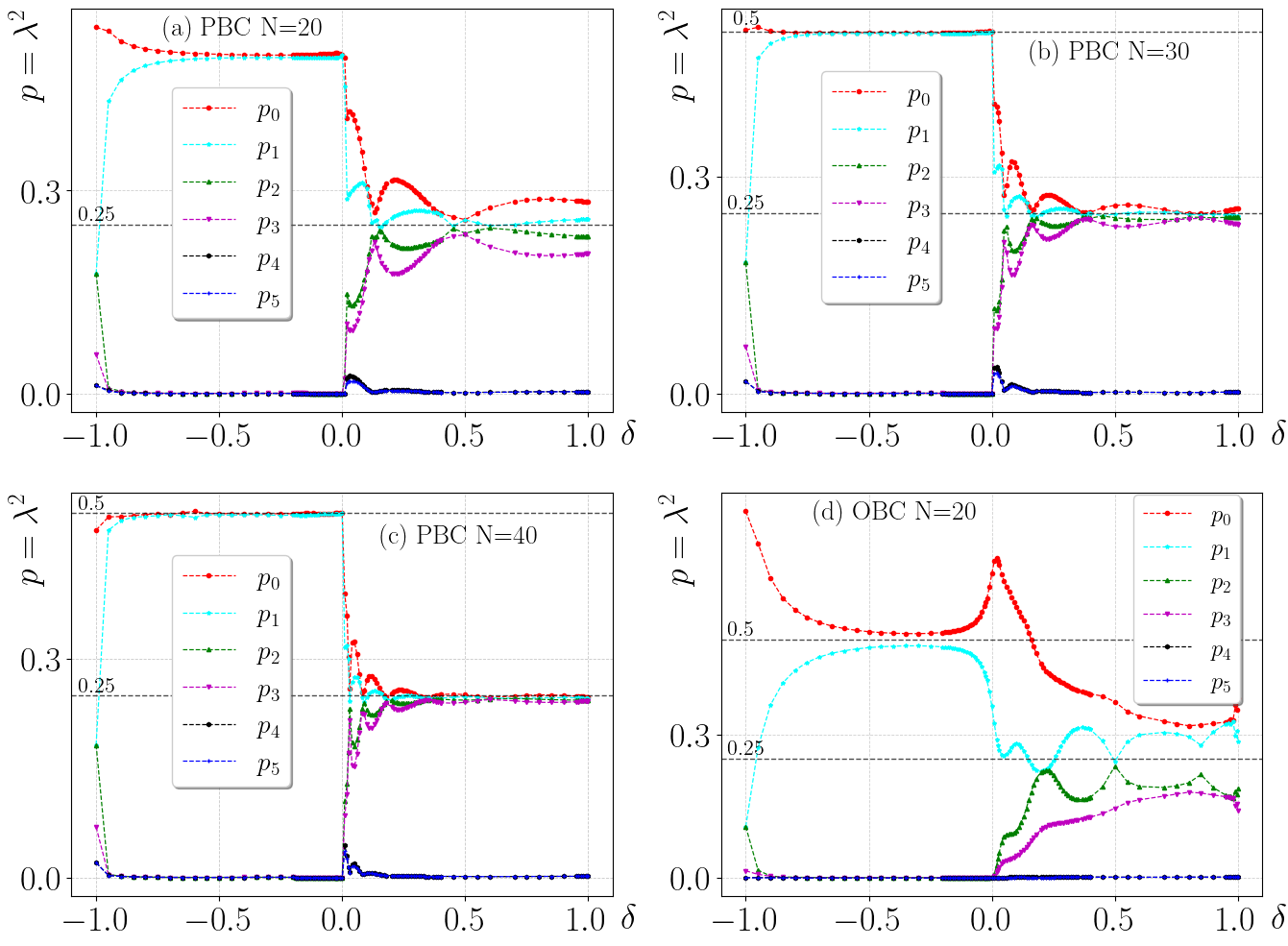}
\caption{
Entanglement structure analysis via Schmidt coefficients for a half-chain bipartition. \textbf{(a)} Dominant Schmidt coefficients $\lambda_\alpha$ as a function of  $\delta$ for $N=20$ under PBC. For $\delta < 0$, two coefficients dominate with $\lambda_1 \approx \lambda_2 \approx 0.707$, corresponding to two-channel bipartite entanglement. For $\delta > 0$, four coefficients become significant ($\lambda_\alpha \approx 0.4-0.5$), indicating four-channel bipartite entanglement characteristic of the incommensurate phase.  \textbf{(b)} same as (a)  for $N=30$. Schmidt coefficients become more evenly distributed, with two-channel regime approaching degeneracy and four-channel regime becoming more balanced. \textbf{(c)} Same as (a) $N=40$. Trend toward degeneracy continues, suggesting thermodynamic limit approaches an ideal multi-channel distribution. \textbf{(d)} Same analysis under OBC for $N=20$. Distribution becomes less balanced, moving further from degeneracy, confirming multi-channel structure as a bulk phenomenon requiring translational invariance. 
}
\label{fig:Es_N}
\end{figure}

\subsection{Schmidt eigenvalues}
\label{subsec:entangstruc}

Concurrently, the concept of entanglement has evolved from a subject of foundational debate into a practical tool for probing the internal structure of quantum matter. For one‑dimensional spin chains, the Schmidt decomposition of the ground state provides a complete characterization of how quantum information is shared between two halves of the system~\cite{Vidal2003a, Li2008}. The number of significant Schmidt coefficients, known as the Schmidt rank, directly reveals the effective entanglement dimensionality. A change from two dominant coefficients to four dominant coefficients signals a fundamental restructuring of the entanglement, which we call an entanglement channel crossover~\cite{DeChiara2012}. The entanglement spectrum and the Schmidt gap (the difference between the largest and second‑largest Schmidt eigenvalues) are sensitive probes of quantum phase transitions, often scaling with critical exponents or reflecting the presence of topological order~\cite{Osterloh2002, Pollmann2010}. Fundamentally, this approach provides a universal way to classify quantum phases without local order parameters~\cite{Amico2008}. Technologically, controlling the number of entanglement channels could enable the design of quantum states with specific information‑processing capabilities, such as distributing quantum information across two versus four parallel pathways for quantum communication or error correction~\cite{Eisert2010, Horodecki2009}.

Therefore,  further understanding can be gained from the distribution of Schmidt eigenvalues (entanglement spectrum), because these encode information about entanglement  and its distribution across different modes~\cite{Li2008}. The entanglement spectrum is defined as the eigenvalues $\lambda_\alpha$ of the reduced density matrix $\rho_\mathrm{A}$ (or their logarithms $\xi_\alpha = -\ln \lambda_\alpha$). It provides a detailed fingerprint that reveals  more information than the single-number entropy~\cite{Li2008, Calabrese2008}. 
In particular,  the entanglement spectrum reflects the underlying organization of quantum correlations and can distinguish between different types of entangled states, even when they share the same total entropy~\cite{Pollmann2010, Poilblanc2010}.

For any bipartition of a pure ground state $|\Psi_0\rangle$ into subsystems L and R, the Schmidt decomposition provides a canonical representation
\begin{equation}
|\Psi_0\rangle = \sum_{\alpha=1}^{\chi} \lambda_\alpha |\alpha_\mathrm{L}\rangle \otimes |\alpha_\mathrm{R}\rangle,
\end{equation}
where $\lambda_\alpha > 0$ are Schmidt coefficients satisfying $\sum_\alpha \lambda_\alpha^2 = 1$, and ${|\alpha_\mathrm{L}\rangle}$, ${|\alpha_\mathrm{R}\rangle}$ are orthonormal bases for each subsystem~\cite{Nielsen2012}. The number of non-zero coefficients $\chi$ is the Schmidt rank, and the entanglement entropy is $S_v = -\sum_\alpha \lambda_\alpha^2 \ln \lambda_\alpha^2$. 

As we discuss below, our analysis shows that the entanglement is entirely decomposable into multiple bipartite channels that coexist and compete within the same ground state~\cite{Denker2023, Ghne2009}.
We note that this will distinguish  our  states from other interesting states.  In particular, we will compare to prototypical Symmetry-Protected Topological (SPT) phases,  like those associated to cluster states, or graph states, that require multipartite entanglement~\cite{Hein2006, Raussendorf2001, Briegel2001}.

The entanglement spectrum analysis reveals a complex hierarchical structure with multiple significant Schmidt values, where the number of dominant eigenvalues directly indicates entanglement complexity~\cite{Li2008, Calabrese2008}. As shown in Fig.~\ref{fig:Es_N}, the Schmidt coefficients reveal  two fundamentally different regimes. For almost the whole range $\delta < 0$ (except close to $\delta=-1$), two Schmidt coefficients dominate, with $\lambda_1 \approx \lambda_2 \approx 0.707$ and corresponding probabilities $\lambda_1^2 \approx \lambda_2^2 \approx 0.5$ (Fig.~\ref{fig:Es_N}a). The dominance of two Schmidt values indicates entanglement primarily carried by two competing bipartite channels, each representing a different way of pairing spins, with the small splitting reflecting that frustration prevents either channel from fully dominating~\cite{White1996, Poilblanc2010}. For a range of $\delta$ smaller but close to zero we showed in previous section that these two channels correspond to the superposition presented in the GHZ-like state, Eq.~\eqref{eq:cat}. For the rest of range, that is from close to $\delta=-1$ to around $\delta=-0.2$ the two channels correspond to a state with non-obvious superpositions of spins.  

Moreover, these two coefficients exhibit a small splitting. Our calculations for  $N=40$  (panel c)  keep this small splitting (see discussion below). This is due to  frustration effects introduced by the cluster three-body  terms which avoid exact degeneracy. This exact degeneracy   would happen for  a SPT phase~\cite{Pollmann2010, Chen2013}. Then  the absence of exact degeneracy confirms the system is not in an SPT phase.

For $\delta > 0$, we observe a more complex scenario (see Fig.~\ref{fig:Es_N}a). Across $\delta=0$ there is a change from two to four dominating Schmidt coefficients, with values $\lambda_\alpha \approx 0.4-0.5$ and corresponding probabilities $\lambda_\alpha^2 \approx 0.2-0.3$. This is signaling the QPT in the system,  but it cannot be linked to the so-called Schmidt transition because the phase for $\delta > 0$ is  actually an incommensurate phase.  This is the entanglement signature of the incommensurate phase: competing interactions generate multiple nearly degenerate ways of partitioning quantum correlations, and the ground state maintains a superposition of all these possibilities~\cite{Papanikolaou2012, Liuke2025}. The hierarchical organization with four dominant values and a tail of smaller ones aligns with characteristic patterns in incommensurate phases, where competing interactions generate complex entanglement distributions~\cite{Poilblanc2010, Dell'Anna2013}. 

Boundary conditions significantly shape the entanglement structure, as revealed by comparing PBC and OBC in Fig.~\ref{fig:Es_N}(d). Under OBC, the distribution becomes more uneven: for $\delta < 0$, the two-channel structure becomes less balanced; for $\delta > 0$, the four-channel structure is less pronounced, and the system moves further from degeneracy. This boundary sensitivity confirms the two or four-channel structure is a bulk phenomenon requiring translational invariance to manifest fully~\cite{Alcaraz1987}. 


The entanglement structure observed under PBC can be conceptually understood within the framework of competing orders~\cite{Sachdev2011, Podolsky2012}. The ground state constitutes a quantum superposition of multiple competing bipartite virtual orders, with entanglement entropy measuring the inherent uncertainty regarding which order would emerge upon measurement. Each dominant Schmidt value corresponds to a distinct way of organizing quantum correlations across the cut—different entanglement patterns that compete due to frustration~\cite{White1996, Poilblanc2010}. The elevated entropy in the $\delta > 0$ regime ($S_v > 2\ln 2$) originates specifically from competition between multiple bipartite channels with comparable weights. When two channels dominate, entropy is low ($\approx \ln 2$); when multiple channels compete with comparable weights, entropy increases substantially, scaling roughly as $\ln(\text{number of significant channels})$~\cite{Eisert2010, Laflorencie2016}. The observed $S_v \approx 1.5$ is precisely consistent with four significant channels having probabilities $\approx 0.25$ each ($\ln 4 \approx 1.386$), plus contributions from the tail of smaller Schmidt values.

The observed entanglement structure definitively rules out several alternative phase interpretations: (i) Not a spin liquid: spin liquids exhibit ground-state degeneracy on non-trivial manifolds and specific entanglement spectrum degeneracies related to anyon content~\cite{Kitaev2006, Balents2010}; our system shows no such degeneracy. (ii) Not an SPT phase: SPT phases exhibit characteristic exact degeneracies in the entanglement spectrum protected by symmetries~\cite{Pollmann2010, Chen2013}; our near-degeneracy is not exact and does not follow SPT selection rules. (iii) Not a Valence-Bond Solid (VBS) phase: VBSs have entanglement spectra with two dominant values corresponding to singlet formation across bonds~\cite{Affleck1987, Majumdar1969};  our $\delta < 0$ regime shows two dominant values, the small splitting and absence of dimerization order distinguish it from true VBS. 

The emergence of four dominant bipartite channels in the $\delta > 0$ regime directly links to the frustration-driven incommensurate phase. Incommensurate order arises when competing interactions produce correlations with a period incommensurate with the lattice spacing~\cite{Papanikolaou2012, White1996}. Such states necessarily involve multiple length scales and complex phase relationships, naturally manifesting as multiple competing bipartite entanglement patterns in the Schmidt decomposition. Each dominant Schmidt value corresponds to a different way of realizing the incommensurate modulation across the entanglement cut, with the superposition encoding quantum uncertainty in the incommensurate order's phase~\cite{Liuke2025, Dell'Anna2013}. 
%

\section{Comments on the role of different terms in the Hamiltonian}
\label{sec:suu}

\begin{figure}[ht]
    \centering
    \includegraphics[width=1.0\textwidth]{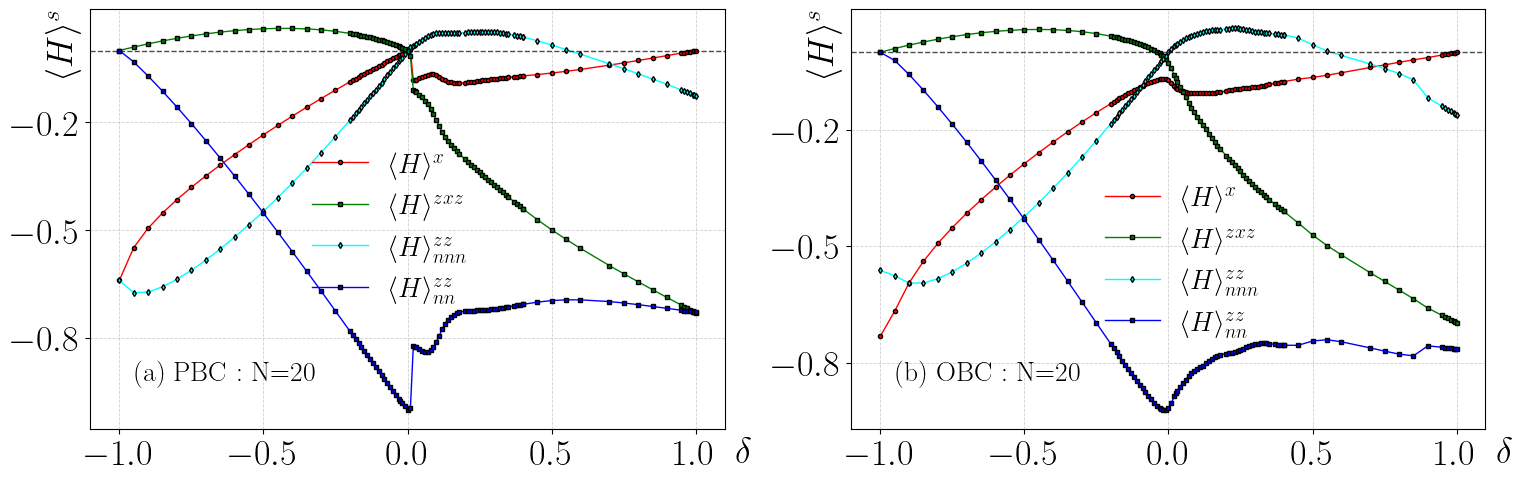}
    \caption{ Comparison of ground-state expectation values of individual Hamiltonian terms $\langle H\rangle^s $ as a function of $\delta$ for a $20$-site spin chain under \textbf{(a)} PBC  and    \textbf{(b)}  OBC. 
Displayed are the four terms multiplied by their coefficients, labeled in the legend as:  
$ \langle H\rangle ^x:-A(\gamma,\delta)\  \langle\sigma_i^x\rangle$ ,  
$\langle  H\rangle ^{xzx} : B(\gamma,\delta) \langle\sigma_{i-1}^z\sigma_i^x\sigma_{i+1}^z\rangle$ ,  
$ \langle H\rangle ^{zz}_{nnn} : \delta\langle\sigma_{i-1}^z\sigma_{i+1}^z\rangle$, and   
$ \langle H\rangle ^{zz}_{nn} : -\frac{\gamma}{2}(\delta+1)\langle\sigma_i^z\sigma_{i+1}^z\rangle$.  
   }
    \label{fig:exp_ham}
\end{figure}

There are further observations that can be obtained from analyzing the expectation values  of the four terms of the Hamiltonian Eq.~\eqref{eq:H_expanded} multiplied by their corresponding coefficients. For PBC the expectation values are site independent. For OBC we average over all combinations compatible with these boundary conditions which are obtained  ranging site index $i$. These are presented in Fig.~\ref{fig:exp_ham}. 

We have not yet discussed  the particular case  $\delta=-1$. This is a peculiar point, and the aim of this paper is not to discuss this in depth. It is nevertheless useful to analyze it briefly to motivate comments about other parts of the phase space. From Fig.~\ref{fig:exp_ham}, we notice that, at this point,  only two terms are different from zero, both for PBC and OBC: NNN term and Ising term $\sigma_i^x$. The difference among   PBC and OBC is that, while for the former both are equal, for the latter $\sigma_i^x$ dominates over NNN. Attending to Fig.~\ref{fig:zz_cor}(b) we see perfect anticorrelations for this case. When $\delta$ departs from -1 slightly, these correlations disappear, and we enter the ferromagnetic phase.  

For $\delta\in(-1,0)$ the cluster three‑body term shows a small contribution: it departs slightly from zero to positive values and returns to zero at $\delta=0$ for PBC. Remarkably, for OBC, it takes  a value slightly below zero at $\delta=0$.  Then, its role on this side is much less important than the other terms.

For $\delta$ slightly above -1 the NNN term decreases slightly and then  increases consistently towards zero for PBC (for OBC it reaches slightly below zero). The $\langle H\rangle ^x$ term increases monotonically towards zero at $\delta=0$ for PBC and well below zero for OBC. One difference between OBC and PBC is that at $\delta=-1$ these two terms are equal for PBC and different for OBC. Second important difference is that they do not reach zero at $\delta=0$ for OBC. 

The most important term in this region, $\delta\in(-1,0)$, is the Ising two-body term. Both for PBC and OBC, this decreases monotonically towards a large negative value which dominate over all other terms when $\delta\to 0^-$. This is the term responsible to establish the ferromagnetic character on this phase and that explains the robust cat states reached when $\delta\to 0^-$. Throughout the other range of parameters in this region the other terms contribute slightly, not sufficiently to destroy order and break symmetry, but enough  to depart from this clear picture at $\delta\to 0^-$. 

At the QPT, close to $\delta=0$, for this finite system, there are striking differences between PBC and OBC. There are finite jumps in the Ising NN, the one-body $\langle\sigma_{i}^x\rangle$ and the cluster three-body terms for PBC. This abrupt jumps are hindered by OBC, where the terms change smoothly. This is in agreement with the first-order QPT well captured by all other quantities describd above. 

In the region $\delta\in(0,1)$ all terms for PBC and OBC are very similar. The Ising NN term stays more or less constant, with slight oscillations due to frustration and incommensurability. The cluster three-body terms becomes large and negative and of the same order as the later for $\delta\to 1$. The  one-body $\langle\sigma_{i}^x\rangle$ term stay more or less constant and of small magnitude, like the NN term. Nevertheless NNN term competes to avoid topological order to be established and keeping the phase incommensurate with the lattice.

\section{Discussion}
\label{sec:Discussion}

Our comprehensive analysis of the Glauber-Ising model reveals rich quantum phases  driven by competing interactions (see Fig. \ref{fig:exp_ham}). Through synergistic application of multiple complementary probes, i.e., spin-spin correlations, static structure factor, Binder cumulant, spectral gap, cat-state fidelity, quantum Fisher information, and entanglement measures, we have established two fundamentally distinct phases separated by a first-order quantum phase transition at  $\delta = 0$.

For \(\delta < 0\), all diagnostics consistently point to a gapped phase with long-range ferromagnetic correlations (gap closes in thermodynamic limit in this region). The spin-spin correlation function \(C_{zz}(r)\) saturates to a distance-independent constant increasing monotonically as \(\delta \to 0^-\), approaching unity at the critical point. This real-space signature is confirmed in Fourier space by the sharp \(q=0\) peak in the static structure factor \(S(q)\). Despite \(\langle \sigma^z_k \rangle = 0\) in finite systems under PBC—a consequence of exact \(\mathbb{Z}_2\) symmetry preservation—the Binder cumulant \(U_z\) approaches the characteristic bimodal value \(2/3\) as \(\delta \to 0^-\), providing unambiguous evidence for spontaneous symmetry breaking in the thermodynamic limit. The spectral gap \(\Delta E_{\text{PBC}}\) exhibits exponential scaling \(\Delta E_{\text{PBC}} \sim e^{-\alpha N}\) with \(\alpha \approx 0.26\), yielding correlation length \(\xi = 1/\alpha \approx 3.8\). This exponential closure definitively signals a first-order QPT driven by macroscopic quantum tunneling. The cat-state analysis reveals that the ground state evolves from a dressed ferromagnetic state  to an essentially perfect symmetric cat state \(|\Phi^+\rangle\) at \(\delta = 0^-\), with fidelity exceeding \(0.999\). The quantum Fisher information \(\mathcal{F}_Q\) saturates the Heisenberg limit \(\mathcal{F}_Q = 4N^2\) as \(\delta \to 0^-\), confirming maximal multipartite entanglement and optimal metrological usefulness. The entanglement structure is characterized by two dominant Schmidt coefficients with a small splitting, indicating two competing bipartite channels. The absence of exact degeneracy rules out SPT order. 

For \(\delta > 0\), the system undergoes a dramatic transformation into a gapped phase (even in thermodynamic limit) with incommensurate correlations—fundamentally distinct from a simple paramagnet. The spin-spin correlation function \(C_{zz}(r)\) decays with distance but exhibits clear oscillatory components with wavelength varying continuously with \(\delta\). This real-space modulation is most clearly captured in the static structure factor \(S(q)\), where the single \(q=0\) peak splits into two symmetric peaks at \(q = \pm q^ *\), with \(q^ *\) evolving in paired peaks  from zero. This peak splitting is the definitive Fourier signature of incommensurate order, reflecting the system's compromise between competing ferromagnetic and antiferromagnetic tendencies. The Binder cumulant \(U_z\) collapses abruptly from \(2/3\) to near-zero values and eventually becomes negative, signaling destruction of the bimodal distribution and emergence of non-Gaussian fluctuations characteristic of the frustrated incommensurate phase. The spectral gap shows no exponential closure; instead, the system exhibits gapped behavior confirmed by logarithmic growth of block entanglement entropy. Cat-state overlaps plummet to negligible values within \(\delta \approx 0.04\), and quantum Fisher information collapses from the Heisenberg limit to near zero. The entanglement structure reveals four dominant Schmidt coefficients with comparable weights, indicating four distinct bipartite entanglement channels contributing substantially to the ground state. This multi-channel structure is the entanglement signature of the incommensurate phase: competing interactions generate multiple nearly-degenerate ways of partitioning quantum correlations. Crucially, despite elevated entanglement entropy exceeding \(2\ln 2\), we found small quantum two-body correlations.

Let us discuss more carefully the first-order  QPT at \(\delta = 0\). We find exponential gap closure for \(\delta < 0\) contrasted with finite gap (in the thermodynamic limit) for \(\delta > 0\) provides unambiguous evidence for the discontinuous nature. All diagnostics show sharp changes across \(\delta = 0\), with finite-size scaling of the Binder cumulant yielding precise crossing at \(\delta = 0 \pm 0.01\). The correlation length \(\xi \approx 3.8\) extracted from gap scaling provides a fundamental length scale governing quantum coherence, independently confirmed by entanglement saturation and the range of \(C_{zz}(r)\). For \(N \gg \xi\), the ground state approaches perfect degeneracy and ideal cat-state fidelity; for \(N \lesssim \xi\), finite-size effects become significant.

The rich physics originates from intricate competition between nearest-neighbor ferromagnetic couplings, next-nearest-neighbor couplings that become antiferromagnetic for \(\delta > 0\), and the three-spin cluster term imposing nonlocal constraints. Frustration prevents simple ordered or disordered states, driving the system toward the incommensurate phase where correlations exhibit continuous tunability with \(\delta\). The three-spin cluster term plays a particularly important role in shaping the entanglement structure, generating correlations beyond simple two-body descriptions and contributing to the multi-channel entanglement observed.

Our incommensurate phase shares features with those found in other frustrated systems such as the ANNNI model~\cite{Selke1988, Bak1982} and frustrated Heisenberg chains~\cite{White1996, Papanikolaou2012}. However, several aspects distinguish our system: the three-spin cluster term generates a richer entanglement structure with four dominant Schmidt channels rather than the typical two~\cite{Poilblanc2010}; continuous tunability of the incommensurate wave vector \(q\) demonstrates high control over modulation period, potentially useful for quantum simulation~\cite{Georgescu2014, Daley2022}; and simultaneous achievement of perfect cat-state fidelity and Heisenberg-limited quantum Fisher information at the critical point highlights the potential of first-order transitions, as the one encountered here, as generators of maximally entangled states~\cite{Ho2019, Stephen2025}.

The emergence of perfect cat states with \(>99.9\%\) fidelity and Heisenberg-limit quantum Fisher information has significant implications for quantum technologies~\cite{Giovannetti2006, Degen2017}. Unlike complex gate sequences or error correction required elsewhere, our system naturally produces maximally entangled states simply by tuning to the phase transition~\cite{Monz2011, Jin2025}. This "criticality-enhanced entanglement generation" could provide a scalable pathway to preparing high-fidelity GHZ states for quantum sensing and metrology~\cite{Zanardi2006, Garbe2020}. The continuous tunability of the incommensurate wave vector also suggests possibilities for quantum simulation of modulated magnetic orders with tailored correlation patterns~\cite{Georgescu2014, Daley2022}. Our findings, validated by DMRG calculations across system sizes up to \(N=40\) with ED benchmarks up to \(N=20\), establish this model as a paradigmatic example of frustration-driven quantum phases and highlight the central role of entanglement in characterizing novel states of quantum matter.

\paragraph*{Experimental feasibility}

The now familiar terms in Hamiltonian Eq.~\eqref{eq:H_expanded}, i.e.,  cluster three-spin term, NN Ising couplings, and NNN  Ising couplings may appear challenging to realize in experiment, but recent advances in quantum simulation platforms have demonstrated that all necessary ingredients occur in different systems, and may be reachable.  In Superconducting devices  non-trivial topology of cluster Ising model  in superconductors has been recently experimentally realized~\cite{2026Tan}. 
In ultracold spinor gases in optical lattices, coherent three-body spin interactions have been directly detected, and the extended Bose-Hubbard model naturally generates effective three-site terms alongside NN and NNN Ising couplings~\cite{Binegar2026}. Also in ultracold atoms, the cluster term can be realized in  the aforementioned optical triangular lattice~\cite{2010Becker}. In superconducting qubit arrays, tunable three-body interactions between flux qubits have been demonstrated using a coupling module that mediates both two-local and three-local interactions by design~\cite{Menke2022}, and three-transmon systems coupled via flux-tunable couplers enable direct realization of multi-qubit interactions with fidelities around 99\%~\cite{Glaser2023}. In Rydberg atom arrays, the facilitation regime (anti-blockade) naturally produces kinetically constrained dynamics where effective three-body interactions emerge, with recent experiments demonstrating controllable three-body interactions where two-body terms are suppressed entirely~\cite{Gambetta2020, Mazza2020}. In trapped-ion systems, controllable three- and four-body spin interactions have been directly observed, and explicit schemes for engineering pure three-body Hamiltonians have been proposed~\cite{Katz2023, Andrade2022}. Together, these experimental capabilities confirm that the full Hamiltonian\eqref{eq:H_expanded} may be implemented in state-of-the-art quantum simulators, making its predicted ground-state properties and phase diagram directly testable.



\section*{Acknowledgements}
We thankfully acknowledge fruitful discussions with Maciej Lewenstein, Utso Bhattacharya and Manuel Gessner. A.P. and M.L.B. acknowledge for support from the project PID2023-152724NA-I00, with funding from MCIU/AEI/10.13039/501100011033 and FSE+, the Severo Ochoa Grant CEX2023-001292-S, Generalitat Valenciana grant CIPROM/2022/66, the Ministry of Economic Affairs and Digital Transformation of the Spanish Government through the QUANTUM ENIA project call - QUANTUM SPAIN project, and by the European Union through the Recovery, Transformation and Resilience Plan - NextGenerationEU within the framework of the Digital Spain 2026 Agenda, and by the CSIC Interdisciplinary Thematic Platform (PTI+) on Quantum Technologies (PTI-QTEP+). Also funding from Horizon Europe EU projects MSCA-SE CaLIGOLA, Project ID: 101086123, and MSCA-DN CaLiForNIA, Project ID: 101119552. These authors gratefully acknowledge the computer resources at Artemisa, funded by the European Union ERDF and Comunitat Valenciana as well as the technical support provided by the Instituto de Fisica Corpuscular, IFIC (CSIC-UV).  
 M.A G-M acknowledges support  from the Ministry for Digital Transformation and of Civil Service of the Spanish Government through the QUANTUM ENIA project call—Quantum Spain project, and by the European Union through the Recovery, Transformation and Resilience Plan—NextGenerationEU within the framework of the Digital Spain 2026 Agenda: also from Projects of MCIN with funding from European Union NextGenerationEU (PRTR-C17.I1) and by Generalitat Valenciana, with reference 20220883 (PerovsQuTe) and COMCUANTICA/007 (QuanTwin).
T.P. gratefully acknowledges the invitation and financial support provided by the Prometeo Grant for two research visits to Valencia and  funding from the Research Council of Finland through Grant No. 359284 (Finnish Quantum Flagship). This work was also supported by the SUPREME project, which is co-funded by the Chips Joint Undertaking under Grant Agreement No. 101286304.

\appendix
\section{Cat state resilience (change)}
\label{sec:appendix}

\begin{figure}
    \centering
    \includegraphics[trim=2cm 0 0 0,width=0.9\linewidth]{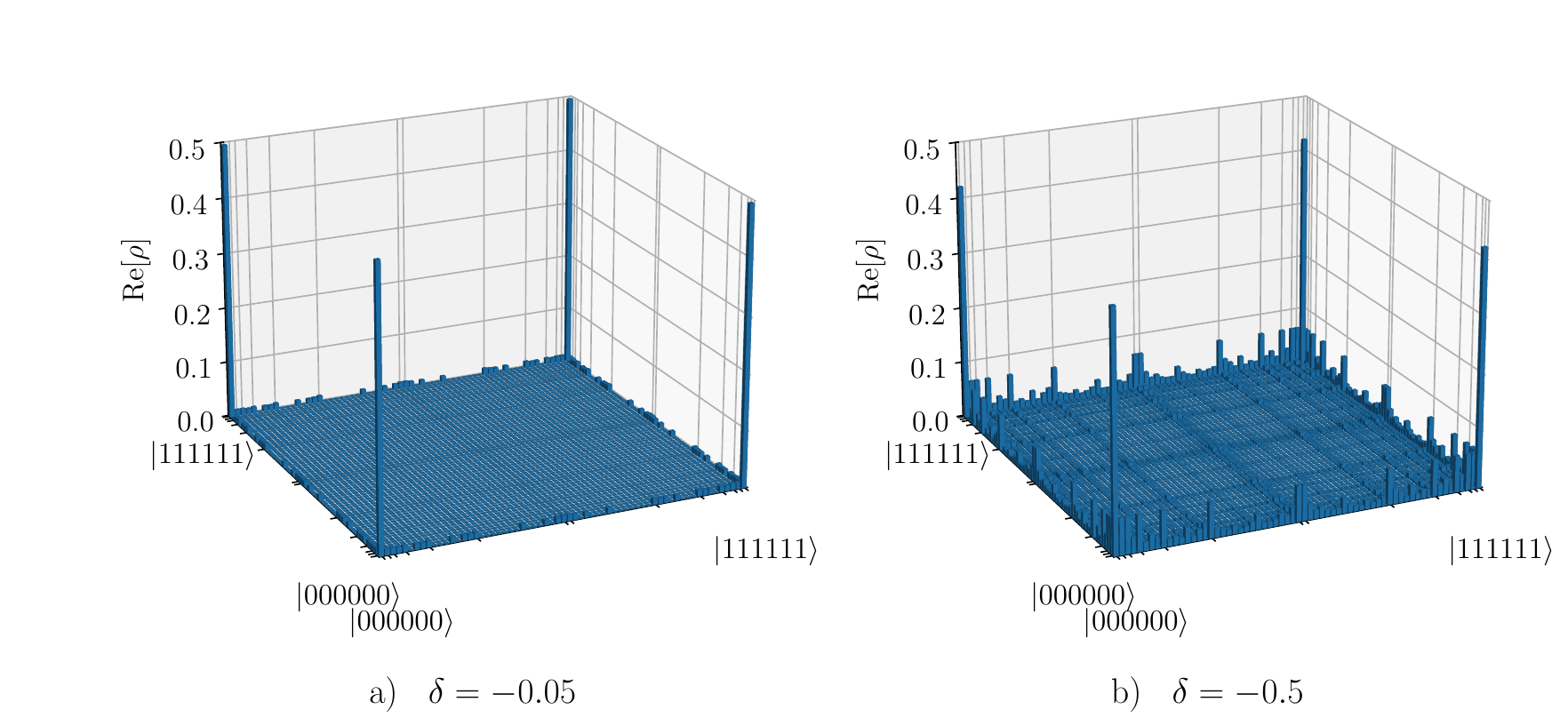}
    \caption{City plot of the density matrix of the ground state for $N=6$ spins, PBC and different values of $\delta$: \textbf{a)} $\delta=-0.05$ and \textbf{b)} $\delta=-0.5$. The vertical bars represents the value of the matrix elements, whose row/column entries are represented in the $x$ and $y$ axis. We only show the $\ket{000000}$ and $\ket{111111}$ states labels, and the others are binary ordering from the first to the last. The grid lines are drawn at the basis states that are at a Hamming distance of 1 from the two cat states. The imaginary parts are not shown since they are exactly null.}
    \label{fig:city_plot}
\end{figure}

In Fig.~\ref{fig:city_plot} we represent the density matrix elements of the ground state, $\rho=|\psi\rangle\langle\psi|$, at different values of $\delta$. Calculations are done using exact diagonalization for $N=6$. This is a large enough chain to illustrate the composition of the ground state for a range of negative values of $\delta$. We present in the figure two exemplary values from our calculations. In this figure, the matrix elements presented are plotted for the basis elements ordered in each axis  as follows. We identify  $|\uparrow\uparrow\cdots\uparrow\rangle$  with $|00\cdots0\rangle$ and we count in binary numbers up to $2^N$, the latter number being  $|11\cdots1\rangle$  that corresponds to $|\downarrow\downarrow\cdots\downarrow\rangle$. When the cat is predominant at $\delta=0$, see Fig.~\ref{fig:Cat_state}, the density matrix displays strong correlations between the two cat states. This we observe for the calculated case at $\delta=-0.05$ shown at panel (a) of  Fig.~\ref{fig:city_plot}. When $\delta$ becomes more negative these correlations dominate but start to become suppressed. New correlations appear between states that are one Hamming distance from the two cat states.  This is illustrated in panel (b) of  Fig.~\ref{fig:city_plot}, where we present the results for $\delta=-0.5$. As mentioned, the peaks that start to contribute to ground state are one spin flip difference from the 0 or $2^N$ states. 

\bibliography{GIM.bib}
\end{document}